# Climate simulations for 1880-2003 with GISS modelE

J. Hansen[1,2], M. Sato[2], R. Ruedy[3], P. Kharecha[2], A. Lacis[1,4], R. Miller[1,5], L. Nazarenko[2], K. Lo[3], G.A. Schmidt[1,4], G. Russell[1], I. Aleinov[2], S. Bauer[2], E. Baum[6], B. Cairns[5], V. Canuto[1], M. Chandler[2], Y. Cheng[3], A. Cohen[6], A. Del Genio[1,4], G. Faluvegi[2], E. Fleming[7], A. Friend[8], T. Hall[1,5], C. Jackman[7], J. Jonas[2], M. Kelley[8], N.Y. Kiang[1], D. Koch[2,9], G. Labow[7], J. Lerner[2], S. Menon[10], T. Novakov[10], V. Oinas[3], Ja. Perlwitz[5], Ju. Perlwitz[2], D. Rind[1,4], A. Romanou[1,4], R. Schmunk[3], D. Shindell[1,4], P. Stone[11], S. Sun[1,11], D. Streets[12], N. Tausnev[3], D. Thresher[4], N. Unger[2], M. Yao[3], S. Zhang[2]

[1]NASA Goddard Institute for Space Studies, 2880 Broadway, New York, New York, USA.
[2]Columbia University Earth Institute, New York, New York, USA.
[3]Sigma Space Partners LLC, New York, New York, USA.
[4]Department of Earth and Environmental Sciences, Columbia University, New York, New York, USA.
[5]Department of Applied Physics and Applied Mathematics, Columbia University, New York, New York, USA.
[6]Clean Air Task Force, Boston, Massachusetts, USA.
[7]NASA Goddard Space Flight Center, Greenbelt, Maryland, USA.
[8]Laboratoire des Sciences du Climat et de l'Environnement, Orme des Merisiers, Gif-sur-Yvette Cedex, France.
[9]Department of Geology, Yale University, New Haven, Connecticut, USA.
[10]Lawrence Berkeley National Laboratory, Berkeley, California, USA.
[11]Massachusetts Institute of Technology, Cambridge, Massachusetts, USA.
[12]Argonne National Laboratory, Argonne, Illinois, USA.

Corresponding author: James Hansen, jhansen@giss.nasa.gov, 1-212-678-5500.

## Abstract

We carry out climate simulations for 1880-2003 with GISS modelE driven by ten measured or estimated climate forcings. An ensemble of climate model runs is carried out for each forcing acting individually and for all forcing mechanisms acting together. We compare side-by-side simulated climate change for each forcing, all forcings, observations, unforced variability among model ensemble members, and, if available, observed variability. Discrepancies between observations and simulations with all forcings are due to model deficiencies, inaccurate or incomplete forcings, and imperfect observations. Although there are notable discrepancies between model and observations, the fidelity is sufficient to encourage use of the model for simulations of future climate change. By using a fixed well-documented model and accurately defining the 1880-2003 forcings, we aim to provide a benchmark against which the effect of improvements in the model, climate forcings, and observations can be tested. Principal model deficiencies include unrealistically weak tropical El Nino-like variability and a poor distribution of sea ice, with too much sea ice in the Northern Hemisphere and too little in the Southern Hemisphere. Greatest uncertainties in the forcings are the temporal and spatial variations of anthropogenic aerosols and their indirect effects on clouds.

## 1. Introduction

Global warming has become apparent in recent years, with the average surface temperature in 2005 about 0.8°C higher than in the late 1800s (*Hansen et al.* 2006a). There is strong evidence that much of this warming is due to human-made climate forcing agents, especially infrared-absorbing (greenhouse) gases (*IPCC* 2001). Concern about human-made climate alterations led to the United Nations Framework Convention on Climate Change (*United Nations* 1992) with the agreed objective "to achieve stabilization of greenhouse gas concentrations in the atmosphere at a level that would prevent dangerous anthropogenic interference with the climate system."

The Earth's climate system has great thermal inertia, yielding a climate response time of at least several decades for changes of atmosphere and surface climate forcing agents (*Hansen et al.* 1984). Thus there is a need to anticipate the nature of anthropogenic climate change and define the level of change constituting dangerous interference with nature. Simulations with global climate models on the century time scale provide a tool for addressing this need. Climate models used for simulations of future climate must be tested by means of simulations of past climate change.

Our present paper describes simulations for 1880-2003 made with GISS atmospheric modelE (*Schmidt et al.* 2006), hereafter *modelE* (2006), specifically model III, the version of modelE "frozen" in mid-2004 for use in the 2007 IPCC assessment. This same model III version of modelE has been documented via a large set of simulations used to investigate the "efficacy" of various climate forcings (*Hansen et al.* 2005a), hereafter *Efficacy* (2005).

*Efficacy* (2005) and the present paper both include use of the same 10 climate forcings. In *Efficacy* (2005) each forcing is based on the fixed 1880-2000 change of the forcing agent, and the mean climate response for years 81-120 is examined to maximize signal/noise. The present paper uses the "transient" (time-dependent) forcings for 1880-2003. These transient simulations are extended to 2100, and in a few cases to 2300, for several GHG scenarios by *Hansen et al.* (2006b, hereafter *Dangerous*

2006) in an analysis of potential "dangerous anthropogenic interference" with climate.

Detailed diagnostics for several of these simulations are available from the repository for IPCC runs (www-pcmdi.llnl.gov/ipcc/about_ipcc.php). Diagnostics for all of these runs, including convenient graphics, are available at data.giss.nasa.gov/modelE/transient.

Sect. 2 defines the climate model and summarizes principal known deficiencies. Sect. 3 defines time-dependent climate forcings and discusses uncertainties. Sect. 4 considers alternative ways of sampling the model's simulated temperature change for comparison with imperfect observations. Sect. 5 compares simulated and observed climate change for 1880-2003, focusing on temperature change but including other climate variables. Sect. 6 summarizes the capabilities and limitations of the current simulations and suggests efforts that are needed to improve future capabilities.

## 2. Climate Model

### 2.1. Atmospheric Model

The atmospheric model employed here is the 20-layer version of GISS *modelE* (2006) with 4°×5° horizontal resolution. This resolution is coarse, but use of second-order moments for numerical differencing improves the effective resolution for the transport of tracers. The model top is at 0.1 hPa. Minimal drag is applied in the stratosphere, as needed for numerical stability, without gravity wave modeling. Stratospheric zonal winds and temperature are generally realistic (Fig. 17 in *Efficacy* 2005), but the polar lower stratosphere is as much as 5-10°C too cold in the winter and the model produces sudden stratospheric warmings at only a quarter of the observed frequency. Model capabilities and limitations are described in *Efficacy* (2005) and *modelE* (2006). Deficiencies are summarized below (Sect. 2.4).

### 2.2. Ocean Representations

We find it instructive to attach the identical atmospheric model to alternative ocean representations. We make calculations with time-dependent 1880-2003 climate forcings with the atmosphere attached to:

(1) Ocean A, which uses observed sea surface temperature (SST) and sea ice (SI). Three 5-member ensembles are run for 1880-2004: (a) SST and SI vary, but climate forcings are fixed at 1880 values, (b) SST, SI, and climate forcings are all time-dependent, (c) SST and forcings vary, but SI is fixed with its 1880 seasonal variation (note that in all cases SI in ocean A is unchanging from 1880 to 1900, because the *Rayner et al.* (2003) sea ice data set begins in 1900).

(2) Ocean B, the Q-flux ocean (*Hansen et al.* 1984; *Russell et al.* 1985), with specified horizontal ocean heat transports inferred from the ocean A control run and diffusive uptake of heat anomalies by the deep ocean. One 5-member ensemble is carried out for 1880-2003 with all climate forcings.

(3) Ocean C, the dynamic ocean model of *Russell et al.* (1995); many simulations are carried out with this model for 1880-2003, including ensembles with each individual climate forcing as well as all forcings acting together. Runs with all forcings have been extended to 2100 and 2300 with several different post-2003 climate forcing scenarios (*Dangerous* 2006). One merit of the *Russell et al.* (1995) ocean model is its computational efficiency. It adds negligible computation time to that for the atmosphere, when the ocean resolution is the same as that for the atmosphere, as is the case here. The ocean model has 13 layers of geometrically increasing thickness, four of these in the top 100 m. The ocean model employs the KPP parameterization for vertical mixing (*Large et al.* 1994) and the Gent-McWilliams parameterization for eddy-induced tracer transports (*Gent et al.* 1995; *Griffies* 1998). The *Russell et al.* (1995) ocean model produces a realistic thermohaline circulation (*Sun and Bleck* 2006), but yields unrealistically weak El Nino-like tropical variability as a result of its coarse resolution.

(4) Ocean D, the *Bleck* (2002) HYCOM ocean model, which uses quasi-Lagrangian potential density as vertical coordinate. Results for this coupled model with all forcings acting at once will be presented elsewhere.

The merits and rationale for organizing the climate change investigation this way, including use of alternative ocean representations with identical atmospheric model and forcings, are discussed by *Hansen et al.* (1997a).

## 2.3. Model Sensitivity

The model has sensitivity 2.7°C for doubled $CO_2$ when coupled to the Q-flux ocean (*Efficacy* 2005), but 2.9°C when coupled to the *Russell et al.* (1995) dynamical ocean. The slightly higher sensitivity with ocean C became apparent when the model run was extended to 1000 years, as the sea ice contribution to climate change became more important relative to other feedbacks as the high latitude ocean temperatures approached equilibrium. The 2.9°C sensitivity corresponds to ~0.7°C per W/m². In the coupled model with the *Russell et al.* (1995) ocean the response to a constant forcing is such that 50% of the equilibrium response is achieved in ~25 years, 75% in ~150 years, and the equilibrium response is approached only after several hundred years. Runs of 1000 years and longer are available upon request. The model's climate sensitivity of 2.7-2.9°C for doubled $CO_2$ is well within the empirical range of 3±1°C for doubled $CO_2$ that has been inferred from paleoclimate evidence (*Hansen et al.* 1984, 1993; *Hoffert and Covey* 1992).

## 2.4. Principal Model Deficiencies

*ModelE* (2006) compares the atmospheric model climatology with observations. Model shortcomings include ~25% regional deficiency of summer stratus cloud cover off the west coast of the continents with resulting excessive absorption of solar radiation by as much as 50 W/m², deficiency in absorbed solar radiation and net radiation over other tropical regions by typically 20 W/m², sea level pressure too high by 4-8 hPa in the winter in the Arctic and 2-4 hPa too low in all seasons in the tropics, ~20% deficiency of rainfall over the Amazon basin, ~25% deficiency in summer cloud cover in the western United States and central Asia with a corresponding ~5°C excessive summer warmth in these regions. In addition to the inaccuracies in the simulated climatology, another shortcoming of the atmospheric model for climate change studies is the absence of a gravity wave representation, as noted above, which may affect the nature of interactions between the troposphere and stratosphere. The stratospheric variability is less than observed, as shown by analysis of the present 20-layer 4°×5° atmospheric model by J. Perlwitz (personal communication). In a 50-year control run Perlwitz finds that the interannual variability of seasonal mean temperature in the stratosphere maximizes in the region of the subpolar jet streams at realistic values, but the model produces only six sudden stratospheric warmings (SSWs) in 50 years, compared with about one every two years in the real world.

The coarse resolution Russell ocean model has realistic overturning rates and inter-ocean transports (*Sun and Bleck* 2006), but tropical SST has less east-west contrast than observed and the model yields only slight El Nino-like variability (Fig. 17 of *Efficacy* 2005). Also the Southern Ocean is too well-mixed near Antarctica (*Liu et al.* 2003), while deep water production in the North Atlantic does not go deep enough, and some deep-water formation occurs in the Sea of Okhotsk region, probably because of unrealistically small freshwater input there in the model III version of modelE. Global sea ice cover is realistic, but this is achieved with too much sea ice in the Northern Hemisphere and too little sea ice in the Southern Hemisphere, and the seasonal cycle of sea ice is too damped with too much ice remaining in the Arctic summer, which may affect the nature and distribution of sea ice climate feedbacks.

Despite these model limitations, in IPCC model inter-comparisons the model used for the simulations reported here, i.e, modelE with the Russell ocean, fares about as well as the typical global model in the verisimilitude of its climatology. Comparisons so far include the ocean's thermohaline circulation (*Sun and Bleck* 2006), the ocean's heat uptake (*Forest et al.* 2006), the atmosphere's annular variability and response to forcings (*Miller et al.* 2006), and radiative forcing calculations (*Collins et al.* 2006). The ability of the GISS model to match climatology, compared with other models, varies from being better than average on some fields (radiation quantities, upper tropospheric temperature) to poorer than average on others (stationary wave activity, sea level pressure).

# 3. Climate Forcings

The climate forcings that drive our simulated climate change arise from changing well-mixed greenhouse gases (GHGs), ozone ($O_3$), stratospheric $H_2O$ from methane ($CH_4$) oxidation, tropospheric aerosols, specifically, sulfates, nitrates, black carbon (BC) and organic carbon (OC), a parameterized indirect effect of aerosols on clouds, volcanic aerosols, solar irradiance, soot effect on snow and ice albedos, and land use changes. Largest forcings on the century time scale are for GHGs and aerosols, including the aerosol indirect effect. Ozone global forcing is significant on the century time scale, and the more uncertain solar forcing may also be important. Volcanic effects are large on shorter time scales, and the clustering of volcanoes contributes to decadal climate variability. The soot effect on snow and ice albedos and land use change are small on global average, but they are large forcings on regional scales.

Global maps of the 1880-2003 changes of these forcings are provided in *Efficacy* (2005). In this section we define the assumed atmospheric, surface or irradiance changes that give rise to the forcings, show the time dependence of global mean forcing for each mechanism, and provide partly subjective estimates of the uncertainties.

We tabulate forcings for several forcing definitions for the sake of analysis and comparison with other investigations. Fi, Fa, Fs, and Fe are, respectively, the instantaneous, adjusted, fixed SST (sea surface temperature), and effective forcings (*Efficacy* 2005). Fi, Fa and Fs are *a priori* forcings. The *a posteriori* forcing Fe is inferred from a long climate simulation, thus accounting in a limited way for the efficacy of each specific forcing mechanism. Fe is based on the 100-year response of global mean temperature, so of course it cannot make different forcing mechanisms equivalent in their regional climate responses. The various forcing mechanisms differ in effectiveness primarily because of their varying locations in latitude or altitude (*Hansen et al.* 1997b, *Ramaswamy et al.* 2001). Even the nominally "well-mixed" GHGs differ in their efficacies, because of spatial gradients in their amounts and the spectral location of absorptions (*Efficacy* 2005).

Fi is the easiest forcing to compute, but in some cases it provides a poor measure of the expected climate response. Fa has been used widely, e.g., by *IPCC* (1996, 2001) and *Hansen et al.* (1997b). Fa is the conventional standard forcing, in which the stratospheric temperature is allowed to adjust to the presence of the forcing agent. Fi and Fa, involving only atmospheric radiation, can be calculated rapidly with precise results for a given model, but it is not practical to compute them for some forcing mechanisms such as the indirect aerosol effect. Fs can be computed with a (fixed SST) global climate model for all forcing mechanisms, but accurate evaluation requires a long model run because of unforced atmospheric variability. Fe, dependent on the simulated response of a coupled atmosphere-ocean climate model, requires even more computer resources for accurate definition because of greater unforced variability in coupled models.

Fi, Fa, Fs and Fe form a sequence that usually should provide successively better predictions of the climate model global response to a given forcing mechanism, because each forcing incorporates further climate feedback mechanisms. For this reason, the forcings are successively more model-dependent, and tabulation of several forcings aids comparison and analysis of climate responses from different models.

### 3.1. Greenhouse gases

**3.1.1. Well-mixed GHGs.** Temporal changes of long-lived GHGs can be approximated as globally uniform. Global mean values of gas amounts, from Table 1 of *Hansen and Sato* (2004) (see also data.giss.nasa.gov/modelforce/ghgases), were obtained from appropriate area weighting of in situ and ice core measurements at specific sites, as described by *Hansen and Sato* (2004). Trace gas measurements are from *Montzka et al.* (1999), as updated at ftp://ftp.cmdl.noaa.gov/hats/Total_Cl_Br. Gas amounts are converted to forcings (Fa), for IPCC and alternative scenarios, using ModelE radiation code, except the minor MPTGs (Montreal Protocol Trace Gases) and OTGs (Other Trace Gases), which use conversion factors provided by *IPCC* (2001). The gas amounts are shown in Fig. 2 of *Dangerous* (2006) and resulting forcings in Fig. 1 here. The 1880-2003 adjusted forcing for well-mixed GHGs is Fa = 2.50 W/m². Efficacy is greater than unity for $CH_4$, $N_2O$ and the CFCs (*Efficacy* 2005), yielding an effective forcing for well-mixed GHGs Fe = 2.72 W/m².

Fig. 1 summarizes climate forcings by well-mixed GHGs and the annual growth of this forcing. The growth rate declined from 5 W/m² per century 25 years ago to 3½ W/m² per century more recently as the growth of MPTGs and $CH_4$ declined.

**3.1.2. Other greenhouse gases.** The principal short-lived, and thus inhomogeneously mixed, anthropogenic greenhouse gas is ozone ($O_3$). $O_3$ change of the past century includes both a long-term tropospheric $O_3$ increase due mainly to human-made changes of $CH_4$, $NO_x$ (nitrogen oxides), CO (carbon monoxide), and VOCs (volatile organic compounds), and $O_3$ depletion (mainly in the stratosphere) in recent decades due to human-made Cl and Br compounds (halogens). The tropospheric historical $O_3$ change in our climate simulation is from a chemistry climate model (*Shindell et al.* 2003) driven by prescribed changes of $O_3$ precursor emissions and climate conditions. Stratospheric $O_3$ change in recent decades is included based on observational analyses of *Randel and Wu* (1999). Some influence of strato-spheric $O_3$ depletion on tropospheric $O_3$ change is included by extrapolating $O_3$ trends in the Antarctic all the way to the surface and reducing $O_3$ growth rates in the Arctic troposphere.

Fig. 2 shows the global mean total $O_3$ versus time, the $O_3$ change as a function of altitude and latitude for the periods 1880-1979 and 1979-1997, and the stratosphere and troposphere $O_3$ changes for these same periods as a function season and latitude. The resulting $O_3$ adjusted forcing, with global average 0.28 W/m² over 1880-2003, is illustrated in Fig. 10b of *Efficacy* (2005).

Future stratospheric $O_3$ may increase as halogens decline in abundance as a result of emission constraints, but the $O_3$ amount will also be affected by climate change. Tropospheric $O_3$ would increase strongly for most *IPCC* (2001) scenarios of $CH_4$ and other $O_3$ precursors (*Gauss et al.* 2003). However, it is possible that efforts to control air pollution and climate change may result in tropospheric $O_3$ levels leveling off or even declining. Given that future $O_3$ changes are highly uncertain and probably not a dominant forcing, we keep $O_3$ in our simulations of the 21st century fixed at the 1997 values (*Hansen et al.* 2002).

The other inhomogeneously mixed anthropogenic GHG included in our climate simulations is $CH_4$-derived stratospheric $H_2O$. Production of stratospheric $H_2O$, based on the two-dimensional model of *Fleming et al.* (1999), is proportional to tropospheric $CH_4$ amount with a two-year lag. As shown in Fig. 9 of *Efficacy* (2005) $CH_4$-derived $H_2O$ increases stratospheric $H_2O$ amount from about 3 ppmv to as much as 6-7 ppmv in the upper stratosphere. Simulated $H_2O$ is in good agreement with observations in the lower stratosphere, which is the region that is important for causing climate forcing. Climate forcing due to $CH_4$-derived $H_2O$ for 1880-2000 is about 0.06 W/m².

Change in the total greenhouse gas effective climate forcing between 1880 and 2003 is Fe ~ 3.0 W/m² (Table 1). Our partly subjective estimate of uncertainty, including imprecision in gas amounts and radiative transfer is ~±15%, i.e., ±0.45 W/m². Comparisons with line-by-line radiation calculations (A. Lacis and V. Oinas, personal communication) suggest that $CO_2$, $CH_4$ and $N_2O$ forcings in the climate model are each accurate within several percent, but the CFC forcing may be 30-40% too large. If that correction is needed, it will reduce our estimated GHG forcing to Fe ~ 2.9 W/m². The documented version of modelE, employed for simulations reported here, in *modelE* (2006), *Efficacy* (2005), and *Dangerous* (2006) has GHG forcings as defined in Table 1 here and Fig. 2 of *Dangerous* (2006).

### 3.2. Aerosols

**3.2.1. Tropospheric aerosols.** Aerosol distributions in our climate model in 2000 are shown in Fig. 3a. All aerosols except sea salt and soil dust are time-variable in the current model, i.e., sulfate, black carbon (BC), organic carbon (OC), and nitrate. The changing geographical distributions of sulfate, BC and OC are from an aerosol-climate model (*Koch* 2001) that uses estimated anthropogenic aerosol emissions based on fuel use statistics and includes temporal changes in fossil fuel use technologies (*Novakov et al.* 2003), but the BC and OC amounts are normalized by time and space independent factors defined below. BC and OC sources are fossil fuels and biomass burning, including agricultural fires that occur mainly in the tropics, and forest fires that occur mainly in Asia and North America. Aerosols from biofuels are not included. OC emissions are taken as proportional to BC emissions, with the OM/BC mass ratio being 4 for fossil fuels and 7.9 for biomass burning (*Liousse et al.* 1996), where it is assumed that the organic matter OM = 1.3xOC. The OC/BC ratios are reduced further, by a small amount, via separate normalization factors for OC and BC defined below. The emission ratios are intended to implicitly account for secondary OC formation (*Koch* 2001). Global aerosol distributions are computed with the transport model for 1850, 1875, 1900, 1925, 1950, 1960, 1970, 1980, 1990, interpolated linearly between these dates, and kept constant after 1990.

Aerosols are approximated as externally mixed for radiative calculations. Absorption by BC is increased a factor of two over that calculated for external mixing to approximate enhancement of absorption that accompanies realistic internal mixing of BC with other aerosol compositions (*Chylek et al.* 1995). BC and OC masses of *Koch* (2001) were multiplied by 1.9 and 1.6, respectively, to obtain best correspondence with multispectral AERONET observations (*Sato et al.* 2003). The GISS model includes the effect of humidity on sulfate, nitrate and OC aerosol sizes (*modelE* 2006), which increases aerosol optical thickness and radiative forcing.

*Andreae and Gelencser* (2006) describe widespread occurrence of "brown carbon," produced especially by biomass burning. Brown carbon is not included as an aerosol per se in our modeling, but it is approximated by the combination of black and organic carbon. Spectral variation of absorption by organic carbon is based on measurements of *Kirschstetter et al.* (2004). Although more detailed treatment of carbonaceous aerosols is desirable, it is difficult to justify that with current measurement limitations (*Novakov et al.* 2005).

Dry nitrate in 1990 is from *Liao et al.* (2004), with nitrate at other times proportional to global population (www.un.org/population). Nitrate aerosol size is taken as similar to the overall aerosol size distribution (*Ten Brink et al.* 1997), with effective radius 0.3 μm and effective variance 0.2. The nitrate aerosol refractive index at 633nm wavelength (1.55 for the dry aerosol) is from *Tang* (1996) and *Tang and Munkelwitz* (1991), with spectral variation the same as for sulfate aerosol.

Fig. 3b shows the 1990 clear-sky and cloudy-sky (i.e., global mean) optical thickness of aerosols in the model. The cloudy-sky aerosol optical thickness, because of higher humidity in cloudy gridboxes, is twice the clear-sky case, but the clear-sky value is appropriate for comparison with observations. Fig. 3b also shows the aerosol adjusted forcing, Fa, and the reduction of downward solar radiation at the surface due to aerosols. The left side of Fig. 3b is the effect of all aerosols present in 2000, while the right side shows the change between 1850 and 2000.

The aerosol effective forcing, i.e., the product Fe = EaFa, varies with aerosol type (Table 2 in *Efficacy* 2005). Fe differs notably from Fa for BC aerosols, as the efficacy of BC is less than 100%. The efficacy depends on the vertical and geographical distribution of the BC, with the reduction of forcing being greater for biomass burning BC (E ~ 60%) than for fossil fuel BC (E ~ 80%) (*Efficacy* 2005). Fig. 3c shows the time dependence of the global mean aerosol optical thickness and effective forcing.

**3.2.2. Aerosol indirect effect.** We use the same aerosol indirect effect on clouds as in *Efficacy* (2005), i.e., a parameterization based on empirical effects of aerosols on cloud droplet number concentration (*Menon and Del Genio* 2006). It is argued in *Efficacy* (2005), based on empirical evidence, that the predominant aerosol indirect effect occurs via cloud cover change, and that the global-mean magnitude of the indirect aerosol forcing, in recent years relative to 1850, is of the order of -1 W/m², with a largely subjective estimate of uncertainty of at least 50%.

Thus, as in *Efficacy* (2005), the scale factor in the indirect effect on cloud cover is chosen to yield a forcing -1 W/m². However, this choice should be thought of as being an approximation for the entire aerosol indirect effect, as we do not explicitly include a cloud albedo effect. The indirect forcing that we employ is smaller than in most models reviewed by *Lohmann and Feichter* (2005), but some recent studies suggest even smaller values, e.g., -0.6 to +0.1 W/m² (*Penner et al.* 2006) and -0.3 to -0.4 W/m² for the albedo effect (*Quaas and Boucher* 2005).

The aerosol indirect effect, as defined by this parameterization, depends on the logarithm of the concentration of soluble aerosols and thus the effect is non-linear, with added aerosols becoming relatively less effective as their number increases. Time-dependent aerosols are anthropogenic sulfates, BC, OC and nitrates, as shown in Fig. 3. Maps of the resulting aerosol indirect forcing are provided in *Efficacy* (2005).

The net 1880-2003 direct aerosol forcing in our transient climate simulations (Table 1) is Fa = -0.38 W/m² and Fe = -0.60 W/m². The total aerosol forcing including the indirect effect is Fe = -1.37 W/m². Empirical data for checking model-based temporal changes of tropospheric aerosol amount, e.g., ice core records (*IPCC* 2001; *Hansen et al.* 2004), are meager. There is a wide spread in aerosol properties inferred from current satellite sensors, but more accurate results are anticipated from future polarization measurements designed to retrieve aerosol and cloud particle properties (*Mishchenko et al.* 2004, 2006). Our largely subjective estimate of the uncertainty in the net aerosol forcing is at least 50%.

**3.2.3. Stratospheric aerosols.** The history of stratospheric aerosol optical thickness that we employ is an update of the tabulation of *Sato et al.* (1993) available at data.giss.nasa.gov/modelforce/strataer. The effective particle radius is ~0.2 μm when the optical depth is small, increasing to ~0.6 μm after the largest volcanoes, as specified on the indicated website and also illustrated by *Hansen et al.* (2002). Aerosols are assumed to have the optical properties of 75% sulfuric acid solution in $H_2O$ (*Palmer and Williams* 1975).

The adjusted forcing by stratospheric aerosols in our model, for aerosols distributed over most of the globe, is (*Efficacy* 2005)

$$Fa \ (W/m^2) \sim -25 \ \tau \ , \tag{1}$$

where τ is the optical thickness at λ = 0.55μm. Because the efficacy of stratospheric aerosol forcing is Ea ~ 91% (*Efficacy* 2005), the effective forcing is

$$Fe \ (W/m^2) \sim -23 \ \tau \ . \tag{2}$$

Published values for the coefficient in (1) range from 20 (*Tett et al.* 2002) to 30 (*Lacis et al.* 1992), with values from different GISS models ranging from 21 to 30. As discussed in *Efficacy* (2005), the result depends on the accuracy of spectral and angular integrations, model vertical resolution, and aerosol distribution. The present result is based on the most accurate of the GISS models, with estimated uncertainty ±15 percent (*Efficacy* 2005).

Satellite observations of the planetary radiation budget perturbation following the 1991 Mount Pinatubo eruption (*Wong et al.* 2004) provide a strong constraint on the aerosol forcing for that volcano (Fig. 11 in *Efficacy 2005*). That comparison suggests that the above relationship between aerosol optical thick-ness and climate forcing is accurate within about 20%.

The stratospheric aerosol forcing becomes more uncertain toward earlier times. We estimate the uncertainty as increasing from ±20% for Pinatubo to ±50% for Krakatau. At intervals between large eruptions prior to the satellite era, when small eruptions could have escaped detection, there was a minimum uncertainty ~0.5 W/m² in the aerosol forcing.

Stratospheric aerosol optical thickness was zero in our climate model control run. Our future control runs will include stratospheric aerosols with $\tau$ ($\lambda = 0.55\mu m$) = 0.0125, the mean visible optical thickness for 1850-2000, with the rationale that this is a better estimate of the long-term mean stratospheric aerosol optical depth than is the use of zero aerosols. We recommend that other researchers include such a mean aerosol amount in control runs used as spin-ups for transient simulations, because the internal ocean temperature will be adjusted to a mean stratospheric aerosol amount. Because the mean aerosol amount is almost 10% of the Krakatau amount, the modeled Krakatau cooling based on a control run with mean aerosol amount is reduced almost 10%, bringing model and observations into better agreement (see Electronic Supplementary Material).

### 3.3. Other Forcings

**3.3.1. Land use.** Changes of land use, especially deforestation that has occurred at middle latitudes and in the tropics, can cause a large regional climate forcing. *Hansen et al.* (1998a) and *Betts* (2001) independently calculated a global forcing of -0.2 W/m² for replacement of today's land use pattern with natural vegetation. Much of the land use change occurred prior to 1880. In *Efficacy* (2005) the time-dependent land use data sets of *Ramankutty and Foley* (1999), illustrated by *Foley et al.* (2005), were found to yield a forcing Fe = -0.09 W/m² for the 1880-1990 change. This forcing may not fully represent land use effects, as there are other land use activities, such as irrigation, that are not included. We do not include the effect of biomass burning burn scars on surface albedo, which *Myhre et al.* (2005) show is a relatively small effect. *Myhre and Myhre* (2003) estimate an uncertainty range from -0.6 to +0.5 W/m² for the land use climate forcing, with positive forcings from irrigation and human plantings, but they conclude that the net land use forcing is probably negative.

We exclude land cover changes occurring as a feedback to climate change, except to the extent they are implicitly included in the *Ramankutty and Foley* (1999) data set. Such land cover changes may have been moderate in the past century, but if the global warming trend of the past few decades continues vegetation feedbacks in the Arctic may be substantial (*Chapin et al.* 2005). This effect can be included in simulations via a dynamic vegetation treatment, but it is not included in our present model.

Our subjective estimate is that the global mean land use forcing for 1880-2000 lies between zero and -0.2 W/m². However, the global value is less relevant than the regional forcing, which can be as much as several W/m², as shown in Fig. 7 of *Efficacy* (2005). The geographical pattern of the climate response is shown in Figs. 18-24 of *Efficacy* (2005) for a fixed forcing and below in Sect. 5 for the transient 1880-2003 forcing.

**3.3.2. Soot effect on snow and ice albedos.** *Clarke and Noone* (1985), from measurements around the Arctic in the early 1980s, showed that soot on snow and ice significantly reduced the albedo for solar radiation. *Hansen and Nazarenko* (2004) estimated that spectrally-integrated albedo changes of 1.5% in the Arctic and 3% in snow-covered Northern Hemisphere land regions would yield a global climate forcing 0.16 W/m² and equilibrium global warming 0.24°C. However, *Grenfell et al.* (2002) and *Sharma et al.* (2004) found smaller soot amounts in more recent measurements, perhaps because of decreased emissions from North America, Europe and Russia, even though emissions from the Far East may have partially replaced those sources (*Koch and Hansen* 2005).

Climate forcing by soot in snow is difficult to simulate well, because albedo change depends sensitively on soot particle structure, how it is mixed in the snow (*Warren and Wiscombe* 1985; *Bohren* 1986), and how much soot is carried away in snowmelt as opposed to being retained near the snow or ice surface. We parameterize snow albedo change as proportional to local BC deposition with a scale factor yielding a conservative estimate of the soot effect, specifically a global forcing Fa ~ 0.05 W/m² in 1990. However, Fe ~ 0.14 W/m² in 1990, because of the high efficacy of snow albedo forcing (*Efficacy* 2005; see also Supplementary Material).

The soot albedo effect is imprecise because of the near absence of accurate albedo measurements and soot in snow inventories and the high efficacy of even a small snow albedo change. Our subjective estimate is that the present soot albedo forcing is probably in the range Fa = 0-0.1 W/m². In some of our simulations there was a programming error that caused this forcing to have an incorrect geographical distribution (see Supplementary Material). The error was corrected for 'all forcings' and snow albedo alone ensembles, but not for simulations illustrated in Figs. 16 because the effect was negligible for the comparisons illustrated in those figures.

**3.3.3. Solar irradiance.** The variations of total solar irradiance in our transient climate simulations submitted to IPCC, shown by the solid curve in Fig. 4, are based on *Lean* (2000). The irradiance changes are largest at ultraviolet wavelengths. The resulting change of climate forcing for 1880-2003 (1880-2000) is Fa = 0.24 W/m² (0.30 W/m²), and Fe = 0.22 W/m² (0.28 W/m²) based on the linear trend, as the efficacy of the solar forcing is Ea ~92% (*Efficacy* 2005). We do not include indirect effects of solar irradiance changes on $O_3$ (*Haigh* 1994; *Hansen et al.* 1997b; *Shindell et al.* 1999; *Tourpali et al.* 2005), which may enhance the direct solar forcing, because the solar forcing itself is moderate in magnitude and uncertain. *Shindell et al.* (2001) conclude that the solar indirect radiative forcing via ozone is small, but there may be dynamical feedbacks that are significant for regional climate change (*Shindell et al.* 2001; *Baldwin and Dunkerton* 2005; *Tourpali et al.* 2005).

*Lean et al.* (2002) call into question the long-term solar irradiance changes, such as those of *Lean* (2000), which have been used in many model studies including our present simulations. The basis for questioning the previously inferred long-term changes is the realization that secular increases in cosmogenic and geomagnetic proxies of solar activity do not necessarily imply equivalent secular trends of solar irradiance. Thus, it is useful to compare the above solar irradiance forcing with a solar irradiance scenario that includes only the well-established Schwabe ~11 year solar cycle, indicated by the dotted curve in Fig. 4. In this alternative solar irradiance forcing history the 1880-2003 forcing based on the linear trend is Fa = 0.10 W/m² and Fe = 0.09 W/m².

The fact that proxies of solar activity do not necessarily imply long-term irradiance change does not mean that long-term solar irradiance change did not occur. Ample evidence for long-term solar change and a link to climate has long been recognized (*Eddy* 1976), and solar models admit the possibility of such change. *Hoyt and Schatten* (1993), on heuristic grounds, argue for solar change at least comparable to that of *Lean* (2000); their inferred solar change is somewhat greater than that of *Lean* (2000) and their secular increase of irradiance begins earlier in the 20th century. At least until precise measurements of irradiance extend over several decades and more comprehensive solar models are available, solar climate forcing is likely to remain highly uncertain.

### 3.4. Summary of Global Forcings

Fig. 5 and Table 1 summarize the time dependence of the forcings that drive our simulated climate change. Effective forcings are shown in Fig. 5, calculated as Fe = EaFa, when Fa is available, and as Fe = EsFs, when Fa is not available, as discussed in *Efficacy* (2005). Use of Fe avoids exaggerating the importance of BC and $O_3$ forcings relative to the well-mixed GHGs and reflective aerosols.

Well-mixed GHGs provide the dominant forcing, which is Fa = 2.50 W/m² and Fe = 2.72 W/m² in 2003 relative to 1880. The total $O_3$ forcing, including tropospheric increase and stratospheric depletion, is Fa = 0.28 W/m² and Fe = 0.23 W/m², as Ea for $O_3$ is 82%. The $CH_4$-derived $H_2O$ forcing is Fs ~ Fe = 0.06 W/m². Thus the total GHG forcing is Fe = 3.0 W/m² in 2003, with $CO_2$ providing about half of the total GHG forcing.

Aerosols, based on our estimates, yield a forcing Fe = −1.37 W/m² in 2003 relative to 1880. Thus the aerosol forcing in our estimate is about half of the GHG forcing, but of opposite sign. The aerosol indirect effect contributes more than half of the net aerosol forcing.

Other effective forcings are solar irradiance (+0.22 W/m² in 2003, a decrease from +0.28 W/m² in 2000), snow albedo (+0.14 W/m²), and land use (-0.09 W/m²).

The sum of all these forcings is Fe ~ Fs ~ 1.90 W/m² in 2003. However, it is more accurate to evaluate the net forcing from the ensemble of simulations carried out with all forcings present at the same time (*Efficacy* 2005), thus accounting for any non-linearity in the combination of forcings and minimizing the effect of noise (unforced variability) in the climate model runs. All forcings acting together yield Fe ~ 1.75 W/m² in 2003.

Uncertainty of the net forcing is dominated by the aerosol forcing, which we suggested above to be uncertain by 50%. In that case, the net forcing is uncertain by ~ 1 W/m², implying uncertainty by about a factor of three for the net forcing. Reduction of this uncertainty requires better data on aerosol direct and indirect forcings.

### 4. Alternative Data Samplings and the Krakatau Problem

Comparisons of simulated climate and observations commonly involve choices that influence how well the model and data appear to agree. Choices of surface temperature data deserve scrutiny, because surface temperature provides the usual measure of long-term 'global warming' as well as a test of climate response to large volcanic eruptions. A number of researchers (e.g., *Harvey and Kaufmann* 2002) have noted that large volcanoes often do not produce the cooling predicted by models. In Supplementary Sect. S1 we examine alternative comparisons of model and observations. Here we briefly summarize principal conclusions from those comparisons.

The model and observations agree more closely when the model is sampled at the locations of observations. The main improvement occurs in the last two decades of the 19th century.

Although it may thus seem best to always appropriately sub-sample the model results, there would be two disadvantages to that approach. First, it makes comparison with other models difficult, because most of these are unlikely to be sampled at the same times and places defined by our specific data sets. Second, it requires additional work and introduces the possibility of error. Because we find that the differences are small in most cases, the global means in this paper are true global means, not a sample at station locations.

Sect. S1 also shows geographical patterns of temperature response after Krakatau and Pinatubo. The model is found to reproduce large scale summer cooling the year after both large volcanic eruptions, and winter cooling with warming in Eastern Europe. Although we cannot fully resolve the issues concerning climate response after large volcanoes, we find the model to be in reasonable accord with observations. This provides support for the model's ability to respond realistically to global forcings.

### 5. Climate Response in Historical Period

Climate model responses to the above forcing mechanisms have been reported in the literature. Nevertheless, side-by-side comparison of responses to each forcing by a single model with documented sensitivity has merit and aids interpretation of the model and real world response to all forcings acting at once. Sect. 5.1 sets the context by showing the response of global mean temperature, planetary radiation balance, and ocean ice cover to all forcings acting at once, and examining the contribution of each forcing to global mean surface temperature change. Sect. 5.2 illustrates the spatial and seasonal distribution of the temperature response to all forcings and individual forcings. Sect. 5.3 examines the effect of all forcings and individual forcings on several other climate variables.

### 5.1. Global Mean Temperature Response versus Time

**5.1.1. Coupled model response to all forcings.** The left side of Fig. 6 compares satellite microwave temperature observations at three atmospheric levels with the coupled climate model response to "all forcings" of Fig. 5. Satellite results in Fig. 6 are from near-nadir observations as analyzed by *Mears et al.* (2003; see also www.ssmi.com/msu/msu_data_description. html), but we compare model results with both *Mears et al.* and *Christy et al.* analyses in Table 2. Although successive versions of the *Christy et al.* (2000) tropospheric analysis have moved from a cooling trend to significant warming, a recent version (5.1) of their analysis (vortex.nsstc.uah.edu/data.msu) has less warming than that in the analysis of *Mears et al.* (2003) (Table 2). Recent assessment of several data sets (*Karl et al.* 2006) concludes that the warming trends of *Mears et al* (2003) are more realistic than those in the analysis of *Christy et al.* (2000).

Fig. 6 includes the microwave lower stratosphere (LS), troposphere/stratosphere (T/S), and middle troposphere (MT) analyses (described as MSU4, MSU3 and MSU2 in prior papers), which are based on near-nadir observations. Use of near-nadir observations yields broad weighting functions, i.e., the derived temperatures refer to thick atmospheric layers, but it avoids increased errors and uncertainty that arises in combining multiple slant-angle data to obtain sharper weighting functions. The LS, T/S and MT levels have weighting functions that peak at altitudes ~15-20 km, ~10 km and ~5 km, respectively (Fig. 2 on *Mears et al.* web site given above). Note that ~15% of the MT signal comes from the stratosphere, despite its description as "middle troposphere".

The simulated global LS warming following the 1991 Mt. Pinatubo eruption agrees closely with observations, which is an improvement over the model results of *Hansen et al.* (2002). The improved response came when the model top was raised from 10 hPa to 0.1 hPa with higher vertical resolution in the stratosphere. The simulated 25-year (1979-2003) trend of global LS temperature (-0.31 °C/decade) agrees well with the *Mears et al.* data (-0.32°C/decade), but not as well with the *Christy et al.* analysis (-0.45°C/decade), as summarized in Table 2.

The simulated T/S temperature trend (+0.10°C/decade) is greater than in the analysis of *Mears et al.* (0.03°C/decade). Temperature change at this atmospheric level is very sensitive to surface temperature. If we replace the coupled model SST with observed SST (ocean A), the discrepancy with the satellite observation largely disappears (Table 2), indeed ocean A yields no warming at that level. As discussed in Sect. 5.3.3, tropical SST variability causes a large variability at the T/S level.

The simulated 25-year MT temperature trend (Fig. 6, Table 2) with all forcings is +0.14°C/decade (+0.15°C/decade for each of the altered aerosol and solar irradiance histories discussed in Sect. 5.4), which is also in good agreement with the observational analysis of *Mears et al.* (+0.13°C/decade) but not with the +0.05°C/decade of *Christy et al.* Observations have greater interannual variability than the model, which is expected as our present coupled model has only slight El Nino-like tropical variability and has unrealistically few sudden stratospheric warmings (Sect. 2.1). The large observed tropospheric fluctuation in 1998, for example, is associated with an unusually strong El Nino.

A sharper lower tropospheric (LT) weighting function can be obtained from linear combination of multiple slant angle microwave observations. Analyses of such LT trends by *Christy et al.* (2000) led to the claim that the lower troposphere was cooling or at least warming much less than surface temperature trends reported by *Jones et al.* (1999) and *Hansen et al.* (2001). *Fu and Johanson* (2005) use linear combinations of near-nadir observations as an alternative approach to obtain tropospheric temperature trends, thus showing that the LT temperature trend of *Christy et al.* (2000) is inconsistent with the near-nadir (MT and LS) data of *Christy et al.* (2000). A recent derivation of LT temperature trends by *Mears and Wentz.* (2005), with an improved diurnal variation of instrumental calibration, yields an LT temperature trend (+0.19°C/decade) that is consistent with our climate simulations (+0.18°C for standard forcings and +0.20°C/decade for the alternative forcings). The recent *Christy et al.* LT trend, from version 5.1 on their web site in September 2005, with their own improved diurnal correction, is +0.12°C/ decade, which is larger than their previous results but less than our model using known forcings and less than the *Mears et al.* analysis of observations (Table 2).

We note that our LS, T/S, MT and LT temperature trends are all obtained using simple vertical weighting functions (*Hansen et al.* 1998b). Resulting global temperature trends for LS, T/S and MT differ little from those obtained with elaborate radiative transfer calculations that include refraction of the microwaves and variable surface emissivity (*Shah and Rind* 1998). However, our calculated LT temperature trend will not account for changes in atmospheric water vapor and surface emissivity, which are substantial for the (slant angle) LT data. Thus our modeled LT change, although a good measure of the model's lower tropospheric temperature change, may not be accurately comparable to the satellite-derived LT 'temperature' trends. For this reason, we emphasize LS and MT data, and, to lesser degree the shorter record of T/S data.

The simulated 1880-2003 global surface temperature change (upper right of Fig. 6), agrees reasonably well with observations, although the 124-year warming based on the linear trend is slightly (~0.1°C) less than observed (Table 2). Two noticeable discrepancies with the temporal variation of observed global surface temperature are the absence of strong cooling following the 1883 Krakatau eruption and the lack of a warm peak at about 1940. We suggested above (Sect. 4.2) that the near-absence of observed cooling after Krakatau may be, at least in part, a problem with the ocean data.

The model's fit with peak warmth near 1940 depends in part on unforced fluctuations, e.g., the runs of *Hansen et al.* (2005b), with nearly identical forcings to those in this paper, appear to agree better with observations. As expected, the runs in which the solar forcing includes only the Schwabe 11-year solar cycle (Fig. 4), available on the GISS web-site and included in Table 2 as AltSol, do not produce peak warmth near 1940. AltSol also differs from the standard "all forcing" scenario in having the sulfate forcing reduced by 50%, thus yielding an 1880-2003 global warming of 0.64°C.

It may be fruitless to search for an external forcing to produce peak warmth around 1940. It is shown below that the observed maximum is due almost entirely to temporary warmth in the Arctic. Such Arctic warmth could be a natural oscillation (*Johannessen et al.* 2004), possibly unforced. Indeed, there are few forcings that would yield warmth largely confined to the Arctic. Candidates might be soot blown to the Arctic from industrial activity at the outset of World War II, or solar forcing of the Arctic Oscillation (*Shindell et al.* 1999; *Tourpali et al.* 2005) that is not captured by our present model. Perhaps a more likely scenario is an unforced ocean dynamical fluctuation with heat transport to the Arctic and positive feedbacks from reduced sea ice.

Fig. 6 also illustrates the planetary energy imbalance, which has grown in recent decades because of the rapid increase of the net climate forcing (Fig. 5) and the ocean's thermal inertia. The simulated imbalance averaged about 0.7 W/m² in the past decade. As discussed in Supplementary Material, our present simulated energy imbalance for the past decade is ~0.02 W/m² less than found by *Hansen et al.* (2005b), because the stratospheric $O_3$ depletion in the latter paper inadvertently was only 5/9 as large as that estimated by *Randel and Wu* (1999).

The simulated decrease of ocean ice cover over the past century, from ~4.25% of the Earth's surface area to ~4%, is only about half as large as suggested by analysis of observations (*Rayner et al.* 2003). Although sea ice observations contain substantial uncertainty, we note that sea ice is more stable in the present model than in previous GISS models. The increased stability of sea ice apparently accounts for the slightly lower sensitivity, 2.7-2.9°C for doubled $CO_2$, of modelE compared with ~3°C for doubled $CO_2$ with the prior GISS model (*Hansen et al.* 2002).

**5.1.2. Effect of alternative oceans.** Fig. 7 and Table 3 show global mean simulations for the same climate forcings as in Fig. 6, but with alternative oceans. Fig. 7a is for ocean A, i.e., specified SST and sea ice that follow the history of *Rayner et al.* (2003). Fig. 7b is for ocean B, i.e., the Q-flux ocean (*Hansen et al.* 1984; *Russell et al.* 1985), with specified horizontal ocean heat transports inferred from the ocean A control run and diffusive uptake of heat anomalies by the deep ocean.

Stratospheric temperature change is similar for ocean A and ocean C, but there is greater year to year variability with ocean A because, unlike the coarse-resolution Russell ocean, the observed SSTs and sea ice capture tropospheric variability such as that due to El Ninos, which in turn affects stratospheric temperature. The resulting net radiation at the top of the atmosphere also has greater year-to-year variability with ocean A, and there is an off-set by a few tenths of 1 W/m² with ocean A, reflecting the fact that the ocean A model with 1880 forcings was out of radiation balance by that amount.

The simulated 1880-2003 global surface temperature change is larger for ocean A than observed, even though ocean A is driven by the SSTs used to compute observed global temperature. We show in the Supplementary Material that this discrepancy seems to be due mainly to calculation of surface air temperature change over the ocean at an altitude of 10 m in model E. This difference can be reduced by using the temperature of the first ocean layer as the "surface" temperature, but that approach has not been the practice in prior climate studies.

Global mean changes obtained with the Q-flux ocean (Fig. 7b) driven by all the climate forcings are similar to those in the coupled dynamical ocean model (Fig. 6). The change in sea ice cover is again much less than the *Rayner et al.* (2003) analysis of observations suggests. The sea ice model, calculated on the atmospheric grid, is the same for all ocean representations (*modelE* 2006). Lack of sufficient sea ice response may be related to under-prediction of seasonal sea ice change in the modelE control runs.

**5.1.3. Response to individual forcings.** Fig. 8 shows the global surface temperature response to individual forcings, and the response to all forcings acting at once. There are nine individual forcings, as opposed to 10 in Fig. 5a, because the reflecting and absorbing (BC) aerosols are included together in the tropospheric aerosol runs. Five 1880-2003 runs were made for each forcing, with the runs started from control run conditions at intervals of 25 years. The control run was within ~0.2 W/m² of radiation balance at the points of experiment initiation, so model drift was small. The effect of model drift was reduced by subtracting the change of each diagnostic quantity for the corresponding year of the control run.

The response to well-mixed greenhouse gas (GHG) forcing is a global warming of ~1.0°C over the period 1880-2003. There is a slow warming of 0.25°C over the first 75 years, and then a rapid approximately linear warming of 0.75°C. The change in the warming rate reflects the jump of GHG growth rates between 1950 and 1975 driven about equally by Montreal Protocol Trace Gases (MPTGs) and an increase of the $CO_2$ growth rate (Fig. 4 of *Hansen and Sato* 2004). A decline in the growth rate of the GHG forcing occurred near 1990 due to halt in growth of MPTGs and slowdown of $CH_4$ growth (*ibid.*; see also Fig. 1 above), and this is reflected in the simulated global response to the well-mixed GHG forcing.

Stratospheric (volcanic) aerosols, despite their brief lifetime (e-folding decay time ~ 1 year), have a multi-decadal effect on simulated temperature because of clustering of volcanoes near the beginning of the 1880-2003 period and from 1963-1991. Thus volcanoes, specifically the minimal activity during 1900-1950 compared with the late 19th century and the period beginning 1963, contribute to the relative global warmth at mid-century, as has been noted previously (*Tett et al.* 1999; *Harvey and Kaufmann* 2002).

The global coolings due to aerosol direct and indirect forc-ings are consistent with the temporal variation of their forcings (Fig. 5). Their combined global cooling reaches ~0.55°C by 2003.

Ozone and solar irradiance changes cause global warming +0.08 and +0.07°C, respectively, over the 124-year period. Land use change causes a cooling –0.05°C.

Global mean surface temperature responses to the forcings by $CH_4$-derived stratospheric $H_2O$ and the soot snow albedo effect are small, consistent with the small forcings. The small forcing for $CH_4$-derived $H_2O$ occurs because, at least in our model, the large increase of middle stratospheric $H_2O$ caused by increasing $CH_4$ does not extend down to the tropopause region, where it would be effective in altering surface temperature. The soot snow albedo forcing is small by assumption in the absence of adequate measurements (Sect. 3.3.2).

Observed global warming, as well as the global warming in the model driven by all forcings, has been nearly constant at about 0.15°C/decade over the past 3-4 decades, except for temporary interruptions by large volcanoes. This high warming rate is maintained in the most recent decade despite a slowdown in the growth rate of climate forcing by well-mixed GHGs (Fig. 1 of this paper and Fig. 4 of *Hansen and Sato* 2004). The warming rate in the model is maintained because, by assumption, tropospheric aerosols stop increasing in 1990. Prior to 1990 increasing aerosols partially counterbalanced the large growth rate of positive forcing by GHGs.

The assumption that global aerosol amount approximately leveled off after 1990 is uncertain, because adequate aerosol observations are not available. However, there is evidence that aerosol amount declined after 1980 in United States and Europe (*Schwikowski et al.* 1999; *Preunkert et al.* 2001; *Liepert and Tegen* 2002), consistent with a leveling off or decline in 'global dimming' (*Wild et al.* 2005). Aerosol emissions probably continued to increase in developing countries such as China and India (*Bond et al.* 2004). An implicit conclusion is that future global warming may depend on how the global aerosol amount continues to evolve (*Andreae et al.* 2005), as well as on the GHG growth rate.

## 5.2. Spatial and Seasonal Temperature Change

We present results for each of 10 climate forcings, comparing these with appropriate standard deviations for determination of significance. This approach results in small figures, yet the figure size is sufficient for intended interpretations, which emphasize planetary scale features in the response. Indeed, the figures are a pithy alternative to detailed tables and discussion. This organization has the merit of consistency, and it provides useful information even for forcings that yield mostly 'insignificant' response.

**5.2.1. Global maps of surface temperature change.** Fig. 9 shows observed and simulated surface temperature change for the full period 1880-2003 and four subperiods. The period since 1950 is frequently studied because of more complete observations in the second half of the century. Breakdown into the segments 1880-1940, 1940-1979 and 1979-2003 captures the period of observed cooling after 1940, and the era of extensive satellite observations beginning in 1979. The maps show the local temperature change based on the linear trend. Thus the global mean temperature change, shown on the upper right corner of each map, often differs from the change in Table 1, which gives the difference of the 5-year running mean at the start and end of the indicated time interval. The linear trend yields a less noisy map. We provide both global mean results for a more complete description.

The lowest row in Fig. 9 shows the standard deviation ($\sigma$) in the control run, specifically $\sigma$ for the temperature change in all periods of the control run of length 124, 54, 61, 40 and 25 years in the second half of the 2000-year control run, by which time the control run was near equilibrium, e.g., within 0.1 W/m$^2$ of radiation balance with space. Nominally, if the absolute value of the change simulated for a given climate forcing (the mean change for a 5-member ensemble of runs) exceeds $\sigma$, the change is significant at about the 95% level. Note that $\sigma$ decreases approximately as the inverse square root of the period, consistent with expected uncertainty in estimating a trend imbedded in random noise.

There is substantial congruence in the spatial response to different global forcings, e.g., greenhouse gases and tropospheric aerosols, even though the forcing distributions differ. This canonical response is shown more clearly in *Efficacy* (2005) and examined quantitatively in Harvey (2004). It does not apply to highly local forcings such as land use and snow albedo.

The well-mixed GHGs by themselves cause a global mean warming of ~1°C, about 50% larger than observed warming, as shown also by the line graphs (Fig. 8). Surface temperature response to GHGs varies a lot spatially. Almost all land areas warm more than 1°C while most ocean areas warm between 0.5 and 1°C. However, the Arctic warms more than 2°C, while the circum-Antarctic ocean warms only about 0.2°C. Large Arctic warming is an expected result of the positive ice/snow albedo feedback. The small response of the circum-Antarctic ocean surface is mainly a result of the inertia due to deep ocean mixing in that region (*Manabe et al.* 1991), although deficient sea ice in the control run may contribute. A larger response is obtained in that region in the *Efficacy* (2005) experiments in which the full 1880-2000 forcing is maintained for 120 years. Cooling or minimal warming in the South Pacific sector of the Southern Ocean near the dateline is also obtained in many other coupled models (*Kim et al.* 2005).

All forcings together yield a global mean warming ~0.1°C less than observed for the full period 1880-2003. The primary location of deficient warming is in Europe and in Eurasia downwind of Europe. Indeed, Fig. 9 shows that the model produces cooling over Europe, where observations show substantial warming. Fig. 9 also indicates that this regional cooling is due to aerosols, for which the direct effect over Europe is about –1°C and the indirect effect about –0.5°C. Land use change contributes a cooling of about –0.5°C, but in a region largely confined to the area of assumed large 20[th] century agricultural development (see also Figs. 7 and 18 in *Efficacy* (2005)). Independent comparison of our aerosol optical depths with AERONET data (*Holben et al.* 2001; *Dubovik et al.* 2002) indicate an excessive aerosol amount in Europe and downwind, as discussed in Sect. 5.4, where we carry out sensitivity experiments with aerosol amounts altered to be more consistent with regional AERONET data. Note in Fig. 9 that a large fraction of observed European warming occurred during 1979-2003, when some observations suggest that aerosols had begun to decline in Europe (*Schwikowski et al.* 1999; *Preunkert et al.* 2001; *Liepert and Tegen* 2002). Our simulations had little aerosol change over Europe in 1979-1990 and no change after 1990, except for nitrates, which *increased* in proportion to global population (Sect. 3.2.1).

**5.2.2. Zonal mean surface temperature change versus time.** Fig. 10 shows the zonal mean surface air temperature ver-sus time relative to the base period 1901-1930. The early base period allows the total temperature change at adjacent latitudes to be compared readily. The period prior to 1900 is not suitable for a base period because of limited spatial coverage of observations.

The upper two panels on the left side of Fig. 10 compare observations from only meteorological stations and the combined station plus SST observations, in both cases using the data as analyzed by *Hansen et al.* (2001). Use of SSTs slightly reduces the analyzed temperature change over the century. Damping of the temperature change is consistent with the longer response time of the ocean, but also could be a consequence of the larger unforced variability of temperature over land. The difference in the two observational analyses in the Antarctic region is a consequence of using temperatures anomalies from the nearest latitudes with observations to define the (1901-1930) base period values for the Antarctic region. The early century anomalies closest to Antarctica are rather different in the two data sets, the meteorological stations showing a warming over the 1900-present period, while the ocean data set has negligible change. Observations are too meager to say which data set is more accurate.

Surface air temperature in the model with observed time-varying SST and all forcings is shown in the third panel of Fig. 10. The fourth panel has observed time-varying SST, SI and forcings. Changing sea ice has a noticeable effect at high latitudes because of the large difference between surface air temperature and ocean temperature that can exist in the presence of a sea ice layer. An additional ensemble of runs driven only by changing SST and SI, with radiative forcings fixed at 1880 values, was run and is available on the GISS web site. The corresponding diagram appears very similar to the fourth diagram in Fig. 10, as forcings have little effect on surface air temperature if the ocean and sea ice are specified.

Oceans B (Q-flux) and C (coupled model, Russell's ocean) yield similar zonal mean surface temperature responses to all forcings. As expected, neither model captures substantial ENSO variability. Simulated 1880-2003 warming is slightly larger than observed in the tropics, but smaller than observed at Northern Hemisphere middle and high latitudes. Cooling due to volcanoes, e.g., after 1963 (Agung), 1982 (El Chichon), and 1991 (Pinatubo), is greater than observed, although the discrepancy is exaggerated in these plots by the deficient long-term warming trend at northern latitudes. Note that the Q-flux model has greater warming in the Southern Ocean, where deep mixing in the dynamical Russell ocean limits the surface warming.

The bottom two panels on the left side of Fig. 10 show the unforced variability in the control run, for a single 124-year period and for the mean of five 124-year periods. The latter mean is the model's noise level for a 5-member ensemble mean, which is an appropriate measure of significance of features in the ensemble responses to forcings shown in other parts of the figure. However, the real world made only a single "run" through the 1880-2003 period, so the noise level in a single period of the control run is also a useful measure. As expected, the unforced variability in a single run is about twice as large as in the ensemble mean.

The well-mixed GHGs and tropospheric aerosols yield responses far larger than unforced variability, the signals being larger in the Northern Hemisphere, especially for aerosols. The multi-decadal response to stratospheric aerosols is apparent, as well as shorter-term responses. Forcings with a global magni-tude of a few tenths of a W/m², such as ozone and solar irradiance, yield noticeable responses, with the response to ozone being mainly in the Northern Hemisphere and the response to solar forcing mainly in the tropics. The weak global forcings, i.e., land use and snow albedo, yield weak responses with expected signs.

**5.2.3. Zonal mean temperature change versus altitude and season.** Fig. 11a shows the zonal mean temperature change versus altitude for 1880-2003 and three subperiods (1880-1940, 1940-1979, 1979-2003). The bottom row is two times the standard deviation (σ) of temperature change among periods of relevant length (125, 61, 40, and 25 years). Because experiment results for each forcing are a 5-run mean, simulated temperature changes exceeding 2σ are statistically significant at >99%.

Well-mixed GHGs are a major cause of tropospheric warming and stratospheric cooling, $CO_2$ being the cause of stratospheric cooling (*Harvey* 2000). Some forcings that yield a weak response at the surface, specifically ozone change and $CH_4$-derived water vapor, yield responses much larger than inferred variability at higher atmospheric levels. The assumed solar irradiance increase yields significant warming in the upper atmosphere due to the large irradiance change at ultraviolet wavelengths (*Lean et al.* 2002) that are absorbed at high levels.

Note that substantial temperature changes in the troposphere are often accompanied by temperature changes of the opposite sign in the stratosphere. A primary mechanism for the stratospheric temperature change is the change in stratospheric water vapor, as illustrated for many forcing mechanisms in Fig. 20 of *Efficacy* (2005). Generally the forcings that warm the troposphere inject more water vapor into the stratosphere, which allows the (optically thin) stratosphere to cool more effectively. Of course, those forcings that alter a direct source of stratospheric heating, such as volcanic aerosols, stratospheric ozone, and solar irradiance, can override this stratospheric effect of the tropospheric climate change. In addition, as the infrared opacity of the atmosphere increases (decreases) the radiative lapse rate increases (decreases), thus tending to increase (decrease) the temperature at levels below (above) the mean level of radiation to space. The mean level of emission to space is at altitude about 6 km, but tropospheric convection tends to spread the temperature anomaly of the lower troposphere throughout the troposphere. Thus there is a tendency for temperature changes to be of opposite sign in the troposphere and stratosphere, for forcings that do not provide a direct heating source within the stratosphere.

Fig. 11b shows the zonal mean surface and lower stratosphere (MSU LS channel) temperature changes as a function of month and latitude. Stratospheric cooling, with maximum at the poles, is caused by well-mixed GHGs (mainly $CO_2$), $O_3$ and $H_2O$. The surface temperature change is dominated by the warming effect of well-mixed GHGs, which is partly balanced by the cooling effect of tropospheric aerosols. The net effect of all forcings is compared with observations in the next subsection.

**5.2.4. Comparisons with observations.** Fig. 12 compares observed temperature change to simulations using the coupled model driven by all forcings of Fig. 5. Fig. 12a is the latitude-altitude change of zonal-mean atmospheric temperature for several periods with readily available observational data: 1958-1998, 1979-1998, and 1987-2003. For the first two periods the data are as graphed by *Hansen et al.* (2002) using radiosonde analyses of *Parker et al.* (1997). For the third period the observational data are for four levels: surface data from the analysis of *Hansen et*

*al.* (2001) and satellite MT, T/S and LS levels from the analysis of *Mears et al.* (2003; see also www.ssmi.com/msu/msu_data_description.html).

The main difference between simulations and radiosonde data is that the switchover from warming to cooling occurs at a lower altitude in the radiosonde data and stratospheric cooling is greater. This difference is of the nature identified by *Sherwood et al.* (2005) as spurious cooling in the radiosonde data that increases with altitude, which can at least partly account for the discrepancy. Agreement is better in the period 1987-2003, which employs satellite and surface data, but the model has 0.1-0.2°C too much tropical upper tropospheric warming, consistent with the excessive warming at the surface in the tropics (Figs. 9 and 10). Interpretation of this apparently excessive tropical warming is provided in Sect. 6.1. Near the North Pole observed warming exceeds that modeled, but the discrepancy there is less than the unforced variability at those latitudes (bottom row of Fig. 11b). In spite of these differences, it seems clear that, in both model and observations, there has been slight cooling at the tropopause level (defined in Fig. 3 of *Efficacy* [2005]) in recent decades, as has been discussed (*Zhou et al.* 2001; *Efficacy* 2005) because of its relevance to stratospheric water vapor trends.

Tropopause height must change in response to tropospheric warming and cooling near and above the tropopause. Fig. 12a shows that the level of zero temperature change occurs beneath the tropopause and the degree of cooling increases with altitude at the tropopause level in observations and model. This implies that the tropopause height increased over these periods. *Santer et al.* (2003, 2005a,b) examined temperature profile and tropopause height changes in climate models and reanalysis data, showing that the tropopause height provides a useful verification that the atmosphere is responding as expected to climate forcings. Fig. 11a shows that several forcing mechanisms contribute to tropospheric warming with temperature change decreasing with altitude at the tropopause: well-mixed GHGs, $O_3$, and $CH_4$-derived $H_2O$, while tropospheric aerosols and stratospheric aerosols have an opposite response that would tend to decrease tropopause height, although the response in polar regions is more complex.

Fig. 12b shows simulated tropopause height change in our coupled model versus time for 1880-2003. We use the World Meteorological Organization definition of the tropopause (*WMO* 1957; *Reichler et al.* 1996), as illustrated for the GISS model in Fig. 3 of *Efficacy* (2005). The simulated 1880-2003 tropopause pressure change is about -3.5 hPa, corresponding to a tropopause height increase of about 150 m. Consistent with findings of *Santer et al.* (2003), well-mixed GHGs and $O_3$ are the main contributors to increasing tropopause height in our model, and aerosols considerably reduce the magnitude of the net height increase.

Fig. 12c compares observed and simulated month-latitude temperature change at three levels observed by satellite. The model cools at the lower stratosphere (LS) level. Month-latitude features are less prominent in the model than in observations, which is expected because (1) a 5-run mean is compared with a single realization, and (2) the model's variability is known to be deficient (see Sect. 2.4). Observed October-January cooling at the South Pole is captured by the model, in which the cooling is a consequence of specified $O_3$ depletion there (Fig. 2c). Observed temperature changes at the North Pole do not exceed the variability among individual runs (Fig. 11, bottom row), suggesting that they could be unforced variability. Observed equa-torial warming in February-April, which is clearer at the T/S level, seems to be weakly simulated by the model.

The T/S level, with weighting function peaking at 250 hPa, is within the troposphere at low latitudes, and thus shows warming at low latitudes in observations and model (Fig. 12c). At the MT level, with weighting function peaking at 500 hPa, warming extends to all latitudes except near the South Pole. Cooling near the South Pole is a consequence of $O_3$ and GHG changes (*Thompson and Solomon* 2002, 2005; *Shindell and Schmidt* 2004). Seasonal variation of the cooling at the South Pole at the LS, T/S and MT levels is captured by the model as well as could be expected given the unforced variability there (Fig. 11b). Associated sea level pressure and wind changes are illustrated below.

### 5.3. Other Climate Variables

Many climate variables from all of the simulations described in this paper are available at data.giss.nasa.gov/modelE/ transient. The runs are organized as summarized in the tables of this paper. Diagnostics can be viewed conveniently via the web site as color maps and graphs, and the data can be downloaded. Figs. 13-15 provide examples for several climate variables for 5-member ensembles of simulations with the coupled atmosphere-ocean model driven by all forcings of Fig. 5 and driven by individual forcings. All maps shown here are for the same period (1900-2003) to allow appropriate intercomparison among different variables. Comparison with observations that are available only for shorter periods is meaningful in many cases, as most aerosol forcing was added after 1950 and most greenhouse forcing after 1970 (Fig. 5). All parameters can be viewed and downloaded for arbitrary periods from our web site, including results for the forcings not included in Figs. 13 and 14 ($CH_4$-derived $H_2O$ and volcanic aerosols) that produced the smallest trends for the 1900-2003 period. We excluded these two forcings from Figs. 13-14 to allow space for maps of two standard deviations.

The bottom row in Figs. 13-14 is the interannual standard deviation of the annual mean in years 1301-1675 of the control run. A forced change exceeding the interannual σ should be apparent to casual observers. The next to bottom row of maps is the standard deviation among 104-year changes in the control run. A simulated change of the 5-run ensemble-mean for a given forcing that exceeds σ is significant at ~95% level, while a change exceeding 2σ is significant at > 99%. Of course the significance of global mean changes far exceeds the significance of local changes. The number on the upper right of these maps is the global mean of the local standard deviation.

**5.3.1. Radiation-related quantities.** The first column of Fig. 13 is the simulated 1900-2003 change of net radiation at the top of the atmosphere based on the local linear trend. Positive radiation is downward. The 1900-2003 linear trend for all forcings operating at once is an increase of ~0.5 W/m² of radiation into the planet. Greenhouse gases cause an increase of ~1 W/m², but the aerosol direct and indirect forcings reduce the planetary radiation imbalance by almost half, as the dominant effect of increasing aerosols and clouds is to increase reflection of solar radiation to space. Incoming net radiation increases in the tropics and Southern Ocean (mainly due to well-mixed GHGs), while it decreases at middle latitudes in the Northern Hemisphere (due to aerosol direct and indirect forcings). The change due to GHGs is an amplification of their normal greenhouse effect, which decreases outgoing radiation in the tropics and increases it at the poles. We note that tropical (20N-20S) net radiation imbalance +1.4 W/m² for the past two decades (a period including about half of the increase in climate forcing for the century) measured by satellite (*Wong et al.* 2006) is consistent with the sign and approximate magnitude of the "all forcing" net radiation change in Fig. 13.

Cloud cover increases in most of the Northern Hemisphere in the climate simulations, on average by more than 1% of the sky cover (Fig. 13 column 2). Cloud cover increase is due mainly to the aerosol indirect effect, which primarily increases low clouds and thus causes surface cooling. Greenhouse gases and resulting global warming, by themselves, lead to a small overall decrease of cloud cover in the GISS model (*Del Genio et al.* 2005). The cloud changes occurring as a climate feedback are distributed in all layers, and their net effect on global surface temperature is near neutral (*Del Genio et al.* 2005). Modeled cloud changes as a function of altitude are available on the GISS web site.

Local modeled cloud cover changes at some locations exceed the local standard deviation (second row from bottom of Fig. 13) for unforced variability of the simulated 104-year change. However, note that σ increases for the shorter period of most observations, e.g., σ is about twice as large for a 25-year period compared to a 100-year period (Fig. 9, bottom row). In addition, changes in cloud observing procedures with time make it difficult to detect reliably cloud changes of the expected magnitudes (*Warren and Hahn* 2003), and the cloud 'observations' in Fig. 13 (*Mitchell et al.* 2004) are in part proxy estimates inferred from observed changes in the amplitude of the diurnal cycle of surface air temperature. Thus little trust can be placed in these 'observed' cloud changes.

Modeled solar radiation (SW, for shortwave) incident on the ground (Fig. 13 column 3) decreases substantially, especially in the Northern Hemisphere. Despite increased cloud cover, reduced SW at the surface is primarily the direct effect of anthropogenic aerosols (compare 2nd and 8th rows in Fig. 13). Reduction is due mainly to strongly absorbing BC aerosols, as the main effect of non-absorbing aerosols is to convert direct downward radiation to diffuse downward radiation. Breakdown by aerosol composition can be seen on the GISS web site for the *Efficacy* (2005) paper (data.giss.nasa.gov/efficacy), which includes 120-year simulations for individual aerosol types. Observational SW data in Fig. 13 (*Gilgen et al.* 1998) are for 1950-1990, which includes more than half of the anthropogenic aerosol increase in our model (Fig. 3c). Observed SW decreases exceed modeled values on average, perhaps in part because observing sites tend to be located near aerosol sources. Observed decrease of SW in the United States is larger than in the simulations, however, analyses based on observed sunshine duration (*Stanhill and Cohen* 2001) for 1900-1985 suggest that longer period changes in the United States were smaller. The large simulated decrease in SW in China is in good agreement with observations for 1961-2000 (*Che et al.* 2005).

Modeled snow and ice cover (Fig. 13 column 4) decreases for all forcings together, with GHGs being the most effective forcing mechanism. Snow increases in regions of deforestation, if land use change is the only forcing, but with 'all forcings' this effect is overridden by global warming. The 'observed' change includes only the sea ice change of *Rayner et al.* (2003), which has about a factor of two greater sea ice loss than simulated by our model. If the *Rayner et al.* (2003) data are approximately correct, the sea ice formulation in the GISS model may be too stable, implying that climate sensitivity in the model III version of modelE may be too small. However, we suggest in the Supplementary Material that much of the large sea ice change in the *Rayner et al.* (2003) data set (Fig. 7a) could be spurious. The large rapid sea ice decrease between 1940 and 1945, e.g., occurs in the Southern Hemisphere, where no warming is evident at that time. Available data on sea ice are inhomogeneous in time, and it is possible, for example, that regions of open water behind the ice edges may be unaccounted for in early sparse observations.

Column 5 in Fig. 13 is the change of the amplitude of the diurnal cycle of surface air temperature. With all climate forcings applied, almost all land regions on Earth are calculated to have a decrease in the amplitude of Ts, by a few tenths of a degree Celsius up to a maximum of about 1°C. GHGs and aerosols (both direct and indirect effects) are principal contributors to damping of the diurnal cycle. The indicated impact of each forcing on the diurnal cycle includes the effect of feedbacks, e.g., increased water vapor and cloud changes. In the case of the well-mixed GHGs the primary mechanism reducing the diurnal cycle is increased water vapor, and thus, since the GHG global warming exceeds actual warming, in a sense the contribution of the GHGs is exaggerated. To the contrary, the impact of aerosols and clouds on the diurnal cycle in Fig. 13 is an understatement of their direct effect, as it includes a negative water vapor contribution. Observed diurnal changes in Fig. 13 are an inhomogeneous data collection (*Mitchell et al.* 2004), but it is clear (*Karl et al.* 1993) that there has been a decrease of the amplitude of the diurnal cycle over most land areas of the globe by several tenths of a degree Celsius, comparable in magnitude to the simulated change.

Pan evaporation is closely associated with radiation variables. Observed pan evaporation decreased in most land areas in the last few decades of the 20th century (*Stanhill and Cohen* 2001; *Roderick and Farquhar* 2002; *Ohmura and Wild* 2002). Fig. 14 shows Penman potential evaporation (*Rind et al.* 1997) over land and actual evaporation over ocean. Pan evaporation can be approximated as 0.7 times potential evaporation (*Roderick and Farquhar* 2002). Simulated reductions of evaporation over most land areas are comparable to observations reported in the above references for the latter half of the 20th century. At almost any location twice the standard deviation (2nd row from bottom of Fig. 14) exceeds the locally simulated change of evaporation, the large variability being associated with large unforced cloud variability. Nevertheless, there is a clear decrease of pan evaporation over land, which is consistent with observations. Fig. 14 indicates that the aerosol direct effect is primarily responsible for reduced potential evaporation over land, but the aerosol indirect effect and land use (in areas of large change, Fig. 7 in *Efficacy* 2005) also reduce potential evaporation over land.

Overall, we conclude that there is good qualitative and semi-quantitative agreement between simulated radiation-related quantities and observations, e.g., with the phenomenon of "global dimming" (*Stanhill and Cohen* 2001; *Liepert* 2002; *Cohen et al.* 2004), the reduction of solar radiation incident on the surface. Our model results indicate that dimming is due primarily to absorbing aerosols and secondarily to the aerosol indirect effect on clouds. The same forcings are principal causes of a decrease in the amplitude of the diurnal cycle of surface air temperature, such a decrease being in accord with observational evidence (*Karl et al.* 1993). The model yields a significant increase of global mean cloud cover, largely due to the aerosol indirect effect, but the local effect does not generally exceed local unforced variability in the model. Large positive global net radiation imbalance simulated for the past decade, which is increasing GHGs, is confirmed by comparison of model results (*Hansen et al.* 2005b; *Delworth et al.* 2005) with observations of ocean heat storage (*Willis et al.* 2004; *Levitus et al.* 2005; *Lyman et al.* 2006). A positive energy imbalance in the tropics is consistent with satellite observations (*Wong et al.* 2006), although it is a challenge to obtain an accuracy in satellite observations that is sufficient to measure the small simulated radiation imbalance.

**5.3.2. Precipitation and run-off.** Well-mixed GHGs by themselves yield large increases of simulated global precipitation and run-off (Fig. 14 columns 2 and 3), with precipitation generally increasing in the tropics and at high latitudes while decreasing in the subtropics, as found in many prior modeling studies (*IPCC* 2001). A tendency for these changes remains when all forcings are applied, but the additional forcings make the hydrologic changes weaker and more variable at middle and high latitudes in the Northern Hemisphere, as direct and indirect aerosol effects tend to reduce precipitation and run-off.

Simulated changes of precipitation and run-off at a given location are usually smaller than unforced variability of the trends among individual model runs (Fig. 14, 2nd row from bottom) and much less than interannual variability (bottom row). This large unforced variability makes it difficult to predict local or regional hydrologic changes. Nevertheless, we notice some consistent tendencies among the present runs and a large number of additional runs with slightly altered forcings (Sect. 6 below and *Efficacy* [2005]) including decreased precipitation in the Southwest United States, the Mediterranean region, and the Sahel, and increased precipitation in the Eastern United States. Precipitation observations compiled by *Mitchell et al.* (2004) have increased precipitation in most areas, although decreases in the Sahel and Mediterranean, with little change in the southwest United States. Note that the *Mitchell et al.* (2004) data, graphically available for arbitrary periods on the GISS web site, have more moderate changes over the past half century (when the data are probably more reliable) and better agreement with the model results.

**5.3.3. Sea level pressure and surface wind.** GHGs have the largest effect on annual-mean sea level pressure and surface zonal wind around Antarctica in our simulations (Fig. 14), exceeding the noise in long-term trends and rivaling interannual variations (bottom two rows, Fig. 14). Ozone and solar irradiance changes reinforce the effect of GHGs, but aerosols work in the opposite sense by cooling low latitudes. Thus surface pressure and zonal wind changes for all forcings are less than for GHGs alone. Observations concur with the net effects, as pressure has decreased over the Antarctic continent and wind speed has increased at most coastal stations (*Turner et al.* 2005), which is a trend toward the positive phase of the Southern Annular Mode (SAM) (*Thompson and Wallace* 2000; *Hartmann et al.* 2000).

*Kindem and Christiansen* (2001) and *Sexton* (2001) used observed stratospheric $O_3$ depletion scenarios in global climate models to show that $O_3$ change yielded stratospheric cooling and a trend toward the positive phase of SAM. *Thompson and Solomon* (2002) showed that $O_3$ changes were consistent with trends of temperature and zonal wind as a function of season and altitude, and that $O_3$ depletion was probably responsible for a slight surface cooling trend in East Antarctica, conclusions supported by climate simulations of *Gillett and Thompson* (2003).

*Fyfe et al.* (1999) and *Kushner et al.* (2001) found that well-mixed GHGs also induce a positive response of the SAM in their models, as upper tropospheric warming and lower stratospheric cooling strengthen the polar vortex in all seasons. This effect is enhanced by minimal warming in the deep-mixing circum-Antarctic Southern Ocean during the transient climate phase. *Cai et al.* (2003) find that stabilizing the GHG amount would weaken the SAM, as the high latitude surface warms and reduces the zonal temperature gradient that drives the subpolar jet stream.

*Shindell and Schmidt* (2004), with a version of the GISS model that includes gravity wave drag and a higher model top, find comparable contributions of $O_3$ and GHGs to surface change, with $O_3$ dominating above the middle troposphere. The larger response to GHGs in our present model is likely a tropospheric effect, as we obtain large low-latitude upper-tropospheric warming (Figs. 11a and 12), consistent with larger-than-observed surface warming in the tropical Pacific Ocean (over most periods, albeit not over 1950-2003, Fig. 9). The larger-than-observed tropical Pacific warming and upper tropospheric warming in the satellite era disappear when we use observed SSTs, as noted in Sect. 5.1.1 and Table 2. Given the large fluctuations of observed tropical Pacific warming (Fig. 9), we suggest that the discrepancy between our ensemble mean result and the real world's single realization in the satellite period (1979-2003), including surface cooling of the Antarctic continent, could be a reflection of real world variability rather than a model flaw. Indeed, despite the lack of ENSO-like variability in the present model, some ensemble members have cooling over East Antarctica in 1979-2003. Quantitative analysis requires a version of the model with more realistic tropical variability.

Observed warming in recent decades (Fig. 9) and warming simulated by a large IPCC ensemble of many climate models in response to increasing GHGs (*Carril et al.* 2005) is concentrated in the Amundsen Sea-Antarctic Peninsula-Western Weddell Sea region, where recent acceleration of outlet glaciers has been noted (*Thomas et al.* 2004; *Cook et al.* 2004). Despite the coarse resolution of our model, the warming pattern that we obtain with "all forcings" is consistent with results of the ensemble of IPCC models (*Carril et al.* 2005).

**5.3.4. Zonal mean quantities.** Fig. 15 shows 5-run ensemble mean changes of zonal mean temperature, water vapor, zonal wind, and stream function produced by each forcing for 1950-2003 and 1979-2003 as a function of altitude and latitude. Control run values of these quantities are shown in *Efficacy* (2005). The standard deviation for each quantity among changes in 54-year and 25-year segments of the control run are shown in the bottom row of Fig. 15.

Conspicuous features in the zonal temperature response were discussed in connection with Figs. 11-12. Water vapor changes, including stratospheric $H_2O$, mainly reflect tropospheric temperature change, except for forcings that directly alter stratospheric temperature. The large stratospheric $H_2O$ change due to $CH_4$-oxidation warms the upper troposphere but barely affects the surface temperature.

Note the large unforced variability of zonal winds at high latitudes. In the Northern Hemisphere only GHGs produce a response exceeding the variability among runs, and only on the 54-year period, not 25 years. A greater zonal-wind response is obtained in the Southern Hemisphere for GHG and $O_3$ forcings, but the effect at the surface exceeds unforced variability only for GHGs on the longer time scale.

The salient change of the meridional stream function caused by "all forcings" is a strengthening of the overturning Brewer-Dobson circulation in the stratosphere. However, response of the Hadley circulation in the lower troposphere is the opposite, i.e., a weakening. These changes are due mainly to the well-mixed GHGs. *Knutson and Manabe* (1995) and *Held and Soden* (2006) show that weakening of mean overturning in the lower troposphere, including the Walker circulation, is due to the increasing proportion of vertical energy flux carried by moist convection as global warming increases. Changes of the stream function for different forcings are more systematic and apparent when viewed for individual seasons as in *Efficacy* (2005), where it is shown that stream function changes are approximately proportional to the effective forcing Fe and in the opposite sense for positive and negative Fe.

**5.4. Alternative Aerosol Forcings**

A primary discrepancy between our simulated global temperature change and observations is the small modeled warming at middle latitudes in the Northern Hemisphere. A likely cause of this deficient warming is the large increase of aerosol optical depth in our model between 1880 and 2003. Causes of large aerosol optical depth in our model include (see Sect. 3.2.1):

(1) Sulfate optical depths are from results of *Koch* (2001) available at the time our "IPCC" runs were initiated. More recent results of *Koch* (personal communication), based on new emission data and more efficient removal including the effect of heterogeneous chemical reactions of sulfate and mineral dust (*Bauer and Koch 2005*), reduce the sulfate optical depth about 50%.

(2) Our aerosol distributions were calculated at coarse temporal intervals (Sect. 3.2), the most recent date being 1990, after which aerosol amount is constant (except nitrate, which increases in proportion to global population). Thus we do not capture post-1990 aerosol decreases in Europe and the United States (see Sect. 5.1), nor aerosol increases in developing countries.

It would be useful to rerun some of our experiments with the best current estimates of global aerosol distributions. However, several significant improvements of our ocean and atmosphere models are also now practical. Thus we plan to repeat the transient climate change experiments with the next documented version of our climate model using the best data then available for aerosols and other transient climate forcings. Here we only make two aerosol sensitivity runs with the present climate model, to provide an indication of how changes in the assumed aerosols affect the simulated climate.

In the aerosol "½ Sulfates" experiment the sulfate changes were reduced by 50%, which also reduced the (negative) AIE by about 18%. In the "½ Sulfates + 2x[Biomass Burning]" experiment the temporal change of BC and OC aerosols from biomass burning was doubled, in addition to the reduction of fossil fuel sulfates. Fig. 16 summarizes the impact of these alternative aerosol histories on the simulated temperature change, comparing the results with observations and the standard "all forcing" model runs.

The reduction of sulfates, which changes the net aerosol 1880-2003 global forcing from −1.37 W/m² to −0.91 W/m², increases global warming from 0.53 to 0.75°C for 1880-2003 and from 0.40 to 0.53°C for 1950-2003. Doubling of the biomass burning aerosol change has little further effect on global temperature (Fig. 16a) as the warming from added BC approximately cancels cooling from added OC.

Both alternative aerosol distributions increase the warm-ing at middle latitudes in the Northern Hemisphere to a realistic level as a result of reduced reflective (sulfate) aerosol optical depth. Given the evidence that our standard aerosol amount was excessive, and does not include post-1990 changes, it is likely that aerosols are a principal factor that needs to be improved in the standard forcings. However, the present alternative distributions should only be viewed as sensitivity studies, as acceptable alternatives need to be obtained from first principles using improved aerosol observations and modeling.

There is notable polar amplification of simulated surface warming in the Northern Hemisphere over the past half century and longer periods, in the cases with aerosol amount reduced to a level that yields realistic mid-latitude warming (Fig. 16c). However, observed polar amplification fluctuates greatly depending on the period considered, which is a reflection of how 'noisy' high latitude temperature is (*Sorteberg et al.* 2005; *Bitz and Goosse* 2006.

The gray band in Fig. 16c is defined by the 1σ variation in the 5-member ensemble of runs for our standard case with all forcings. As would be expected, for some periods and latitudes the observations fall outside the 1σ range, although almost all cases are within 2σ. Note the great variability of observed high latitude temperature change depending on the period considered and the large unforced variability of the polar temperatures among model runs. Given that the real world represents a single realization, the large unforced variability at high latitudes indicates that observed climate changes at high latitudes should be interpreted with caution.

## 6. Summary: Model and Data Limitations and Potential

We summarize model limitations and capabilities (Sect. 6.1) and briefly discuss (Sect. 6.2) why, despite the model limitations, extension of model runs into the future can provide meaningful results. In Sect. 6.3 we summarize lessons learned that may be helpful for future climate simulations.

### 6.1. Global Climate Change: 1880-2003

Our climate model, driven by all of the estimated forcings, simulates observed global mean temperature change over the period 1880-2003 reasonably well. The results fit observations better if tropospheric aerosol change is smaller over Europe than it is in our standard 'all forcing' run. There are independent reasons to believe that a reduced aerosol change there is more realistic, as discussed in Sect. 5.4.

The match of simulated and observed global temperature curves is not an indication that knowledge of climate sensitivity and the mechanisms causing climate change is as accurate as suggested by that fit. An equally good match to observations probably could be obtained from a model with larger sensitivity (than 2.9°C for doubled $CO_2$) and smaller net forcing, or a model with smaller sensitivity and larger forcing. Indeed, although detailed temporal and spatial patterns of climate change, in principle, may allow extraction of inferences about both climate sensitivity and forcings, in practice the large unforced variability of climate, uncertainties in observed climate change, poor knowledge of aerosol forcings, and imprecision in the model-generated response to forcings make it difficult to place useful quantitative limits on climate sensitivity from observed climate change of the past century.

On the other hand, paleoclimate evidence of climate change between periods with well-known boundary conditions (forcings) yields a climate sensitivity 3±1°C for doubled $CO_2$ (*Hansen et al.* 1993), so our model sensitivity of 2.9°C for doubled

$CO_2$ is reasonable. The realistic response of this model to the accurately known short-term forcing by Pinatubo volcanic aerosols and the realistic rate of simulated global warming in the past 3-4 decades, when greenhouse gases increased so rapidly that they dominated over other estimated forcings (Fig. 5), suggest that approximate agreement between observed and simulated global temperature is not fortuitous.

The largest discrepancy in simulated 1880-2003 surface temperature change, other than deficient warming in Eurasia, is warming of the tropical Pacific Ocean that exceeds observations (Fig. 9). *Cane et al.* (1997) suggest that reduced warming in that region may occur with global warming due to increased frequency or intensity of La Ninas, but among all models there is a slight tendency toward increased El Ninos with global warming (*Collins et al.* 2005). Our present model would not be able to capture a change in ENSO variability, whether forced or unforced. However, note that the existence of a discrepancy in Pacific Ocean warming depends on the period considered. For example, our modeled tropical Pacific warming is larger than observed for 1880-2003, but not for 1950-2003 (Figs. 9 and 16c). We conclude that observed variability in tropical surface temperature trends is a strong function of ENSO variability, and the differences between observations and simulations in this region may be due simply to unforced variability.

The greatest apparent discrepancy with observations of temperature change versus altitude is the large simulated warming in the tropical upper troposphere. On the 50-year time scale there is no clear evidence that modeled upper tropospheric warming exceeds observations, given increasing evidence that deficient warming there in radiosonde records may be a figment of inhomogeneities in the radiosonde data sources (*Sherwood et al.* 2005; *Free and Seidel* 2005; *Santer et al.* 2005b; *Karl et al.* 2006). However, on shorter time scales our modeled upper tropospheric warming trend is excessive, as shown most convincingly by MSU T/S data for 1987-2003. But this excessive upper tropospheric warming disappears when we use ocean A (observed SSTs) instead of the fully coupled atmosphere-ocean model (complete ocean A results are available at data.giss.nasa.gov/modelE/transient). Thus the excessive upper tropospheric warming is a reflection of larger than observed warming of the tropical Pacific Ocean in the coupled model. Real world trends toward La Nina conditions for 1987-2003, toward El Nino conditions for 1950-2003, and slightly toward La Nina conditions for 1880-2003 (Fig. 9) are within the range of Southern Oscillation variability. There is no reason to expect our model to capture this specific ENSO variability, and there is no inconsistency between upper tropospheric and surface temperature changes. Thus we anticipate that magnified warming in the tropical upper troposphere will become apparent as the tropical ocean continues to warm in response to increasing GHGs.

In summary, simulated climate change for the past century does not agree in detail with observations, nor would it be expected to agree, given unforced climate variability, uncertainty in climate forcings, and current model limitations. But in a broad sense our climate model does a credible job of simulating observed global temperature change in response to short timescale (volcanic aerosol) as well as century time-scale forcings. This model capability provides sufficient reason to examine the model for information about large-scale regional climate effects of practical importance and to extend the climate simulations to investigate potential global consequences of alternative climate forcing scenarios.

## 6.2. Extended Simulations

Given the large uncertainty in the net climate forcing during the past century (Sect. 3.4), due mainly to aerosols, it may be questioned whether useful accuracy can be achieved in simulations of the future. However, GHG climate forcing is expected to be dominant over aerosol forcing in the future; indeed, it seems likely that at least a moderate decrease of aerosol amount will occur (*Andreae et al*. 2005; *Smith et al*. 2005). Therefore, given a specified GHG scenario, with large increases of GHGs as projected by *IPCC* (2001), the relative uncertainty in the net forcing is reasonably small.

Thus, continuations of some of our present 1880-2003 simulations into the future seem justified. We have extended the 1880-2003 "all forcing" simulations of this paper to 2100, and in a few cases to 2300 using fixed 2003 forcings, several IPCC GHG scenarios, and the "alternative" and "2°C" scenarios of *Hansen and Sato* (2004). These extended simulations are described in *Dangerous* (2006).

## 6.3. Lessons Learned

Our control run had negligible stratospheric aerosols, representing only "background" conditions far removed from the injection of aerosols by large volcanoes. If volcanic aerosols are to be included as a forcing in experiment runs, the use of ocean initial conditions from a control run without volcanic aerosols yields misleading results, especially for ocean heat storage.

In Supplement Sect. S2 we suggest that inclusion of a mean aerosol amount in the control run significantly improves realism of results. We now include in our control runs a mean stratospheric aerosol optical thickness $\tau = 0.0125$ at 0.55 μm wavelength, which is the 1850-2000 mean value of the *Sato et al*. (1993) aerosol climatology. Other researchers may wish to consider including this mean stratospheric aerosol amount in their control runs.

Other "lessons learned" concern methods to minimize the effect of model drift (S3) and the effect of mixed definitions of surface air temperature (S3). We also describe minor errors in ozone forcing (S4) and the forcing due to black carbon deposition on snow (S5). The efficacy of the BC snow albedo effect is found to be 2.7 after the programming error is corrected, rather than 1.7 that was reported in *Efficacy* (2005), where the same programming error was included.

**Acknowledgments.** Data that we use for recent greenhouse gas amounts are from the NOAA Earth System Research Laboratory, Global Monitoring Division, where we are particularly indebted to Ed Dlugokencky, Steve Montzka, and Jim Elkins for up-to-date data. We thank Ellen Baum, Tom Boden, Curt Covey, Oleg Dubovik, Hans Gilgen, Danny Harvey, Brent Holben, Phil Jones, John Lanzante, Judith Lean, Forrest Mims, Bill Randel, Eric Rignot for data and helpful suggestions, and Darnell Cain for technical assistance. Research support from Hal Harvey of the Hewlett Foundation, Gerry Lenfest, and NASA Earth Science Research Division managers Jack Kaye, Don Anderson, Waleed Abdalati, Phil DeCola, Tsengdar Lee, and Eric Lindstrom is gratefully acknowledged.

**Table 1.** Climate forcings (1880-2003) used to drive our climate simulations and surface air temperature changes (based on 5-year running mean) obtained for several periods. Instantaneous (Fi), adjusted (Fa), fixed SST (Fs), and effective (Fe) forcings are defined in *Efficacy* (2005).

| Forcing Agent | Run Name | Forcing [1880-2003] | | | | | $\Delta T_s$ change [Year to 2003] | | | |
|---|---|---|---|---|---|---|---|---|---|---|
| | | Fi | Fa | Fs | Fe | | 1880 | 1900 | 1950 | 1979 |
| Well-Mixed GHGs | E3WMGo | 2.62 | 2.50 | 2.65 | 2.72 | | 0.96 | 0.93 | 0.74 | 0.43 |
| Stratospheric H$_2$O | E3OXo | — | — | 0.06 | 0.05 | | 0.03 | 0.01 | 0.05 | 0.00 |
| Ozone | E3O3o | 0.44 | 0.28 | 0.26 | 0.23 | | 0.08 | 0.05 | 0.00 | -0.01 |
| Land Use | E3LUo | — | — | -0.09 | -0.09 | | -0.05 | -0.07 | -0.04 | -0.02 |
| Snow Albedo | E3SNA2o | 0.05 | 0.05 | 0.14 | 0.14 | | 0.03 | 0.00 | 0.02 | -0.01 |
| Solar Irradiance | E3SOo | 0.23 | 0.24 | 0.23 | 0.22 | | 0.07 | 0.07 | 0.01 | 0.02 |
| Stratospheric Aerosols | E3SAo | 0.00 | 0.00 | 0.00 | 0.00 | | -0.08 | -0.03 | -0.06 | 0.04 |
| Trop. Aerosols, Direct | E3TADo | -0.41 | -0.38 | -0.52 | -0.60 | | -0.28 | -0.23 | -0.18 | -0.10 |
| AIE:CldCov | E3TAIo-E3TADo | — | — | -0.87 | -0.77 | | -0.27 | -0.29 | -0.14 | -0.05 |
| | | | | | | | | | | |
| Sum of Above | No Run | — | — | 1.86 | 1.90 | | 0.49 | 0.44 | 0.40 | 0.30 |
| All Forcings at Once | E3f8yo | — | — | 1.77 | 1.75 | | 0.53 | 0.61 | 0.44 | 0.29 |

**Table 2.** Observed and simulated temperature change based on linear trend for indicated atmospheric levels and periods, with standard deviations from 5-member ensembles.

| *(a) Temperature Trend (°C/decade)* | | | | | | | | |
|---|---|---|---|---|---|---|---|---|
| | | **Observations** | | **Model** | | | | |
| **Level** | **Period** | **Mears** | **Christy** | **Standard Forcings** | | **AltAer1** | **AltAer2** | **AltSol** |
| | | | | **Ocean A** | **Ocean C** | **Ocean C** | **Ocean C** | **Ocean C** |
| MSU LS | 1979-2003 | -0.32 | -0.45 | -0.30 ± 0.01 | -0.31 ± 0.01 | -0.301±0.009 | -0.305±0.010 | -0.303±0.007 |
| MSU T/S | 1987-2003 | 0.03 | NA | -0.006 ± 0.02 | 0.10 ± 0.04 | 0.102±0.036 | 0.103±0.016 | 0.113±0.025 |
| MSU MT | 1979-2003 | 0.13 | 0.05 | 0.15 ± 0.01 | 0.14 ± 0.01 | 0.154±0.018 | 0.148±0.021 | 0.151±0.015 |
| MSU LT | 1979-2003 | 0.19 | 0.12 | 0.20 ± 0.01 | 0.18 ± 0.01 | 0.197±0.021 | 0.200±0.025 | 0.197±0.017 |
| | | **GISS** | **Jones** | | | | | |
| Surface | 1979-2003 | 0.16 | 0.18/0.18 | 0.148±0.014 | 0.136±0.011 | 0.147±0.017 | 0.160±0.028 | 0.153±0.017 |
| *(b) Temperature Change (°C)* | | | | | | | | |
| | | **GISS** | **Jones** | | | | | |
| Surface | 1880-2003 | 0.60 | 0.73/0.73 | 0.697±0.005 | 0.528±0.044 | 0.749±0.035 | 0.712±0.038 | 0.639±0.019 |
| Surface | 1900-2003 | 0.57 | 0.69/0.69 | 0.719±0.007 | 0.460±0.036 | 0.641±0.034 | 0.623±0.030 | 0.554±0.036 |
| Surface | 1950-2003 | 0.52 | 0.52/0.56 | 0.507±0.026 | 0.402±0.038 | 0.526±0.036 | 0.540±0.075 | 0.492±0.044 |
| Surface | 1979-2003 | 0.38 | 0.43/0.44 | 0.355±0.034 | 0.325±0.027 | 0.354±0.041 | 0.383±0.067 | 0.366±0.040 |

Mears and Christy's observations are for 1979 (or 1987) through August 2005. Jones data had HadCRUT2/HadCRUT2v. HadCRUT2 is combined land and marine (SST from the Hadley Centre of the UK Meteorological Office; see Rayner *et al.* [2003] for details) temperature anomalies on a 5°×5° grid-box basis. HadCRUT2v is a variance adjusted version of HadCRUT2.

**Table 3.** Global surface temperature changes (from 5-year running mean) obtained with alternative ocean representations, specifically Ocean A (observed ocean surface conditions) and Ocean B (Q-flux ocean).

| **Forcing Agent** | **Run Name** | **Forcing [1880-2003]** | | | | **ΔT$_s$ change [Year to 2003]** | | | |
|---|---|---|---|---|---|---|---|---|---|
| | | **Fi** | **Fa** | **Fs** | **Fe** | **1880** | **1900** | **1950** | **1979** |
| *Observed Ocean Model* | | | | | | | | | |
| SST and Sea Ice | E2OCN | 0.00 | 0.00 | 0.00 | 0.00 | 0.57 | 0.53 | 0.42 | 0.28 |
| SST, Ice & All Forcings | E3OCNf8 | — | — | 1.77 | 1.75 | 0.67 | 0.66 | 0.51 | 0.31 |
| SST, All Forcings | E3SSTf8 | — | — | 1.77 | 1.75 | 0.62 | 0.56 | 0.45 | 0.30 |
| *Q-Flux Model* | | | | | | | | | |
| All Forcings | E3f8qd | — | — | 1.77 | 1.75 | 0.56 | 0.66 | 0.43 | 0.33 |

**Fig. 1.** Climate forcing (W/m²) and its annual changes (W/m² per year) for observed 1880-2003 greenhouse gas changes as tabulated by *Hansen and Sato* (2004). MPTGs and OTGs are Montreal Protocol trace gases and other trace gases (*Hansen and Sato* 2004). Forcings are conventional adjusted forcings, Fa, relative to their value in 1880.

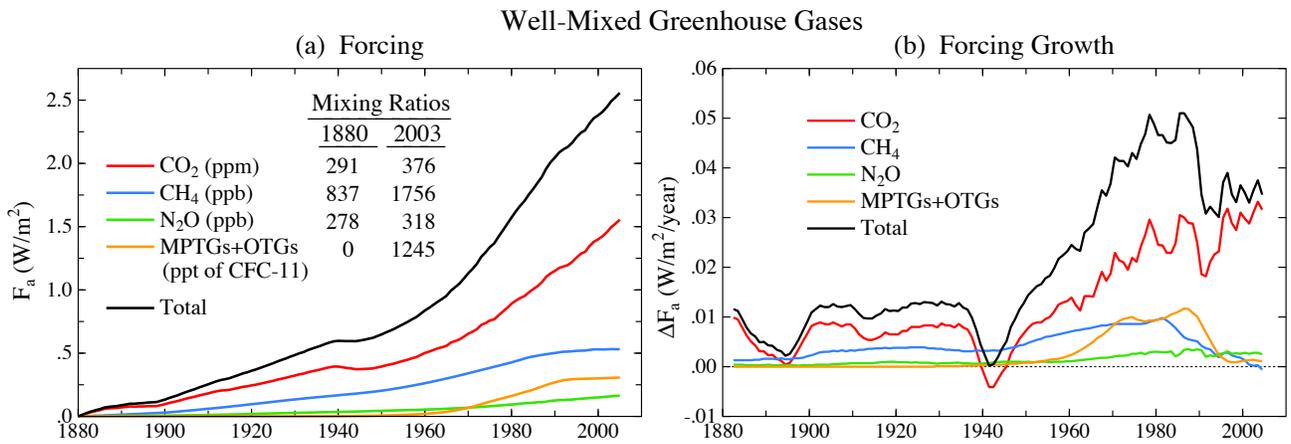

**Fig. 2.** Ozone change in our simulations: (a) global $O_3$ versus time, (b) $O_3$ change versus latitude and altitude (in hPa of pressure) for periods 1880-1979 and 1979-1997, and (c) $O_3$ change in the stratosphere and troposphere versus month and latitude for 1880-1979 and 1979-1997.

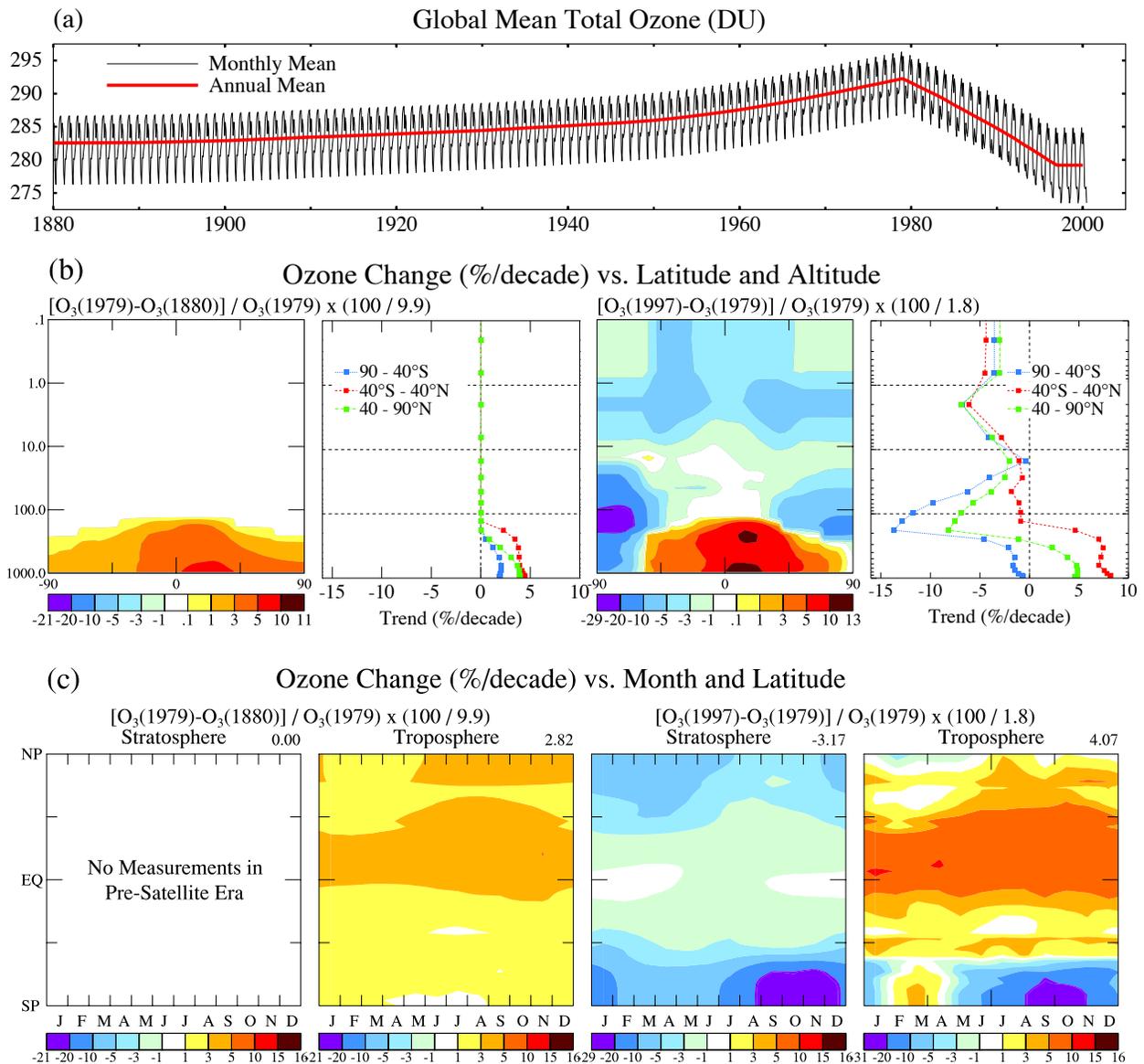

**Fig. 3.** (a) Aerosol clear-sky optical thickness in 2000 in the Model III version of ModelE. Global means are in the upper right corners. (b) Clear-sky and all-sky tropospheric aerosol optical thickness, the resulting adjusted forcing, and the effect on downward solar radiation at the Earth's surface. Left side is the effect of all aerosols in 2000 relative to no aerosols, and right side is the change from 1850 to 2000. (c) Time dependence of aerosol optical thickness (left) and effective forcing Fe = EaFa (right). Effective forcing, Fe, is employed to avoid exaggerating the importance of BC and O$_3$ forcings relative to well-mixed GHGs and reflective aerosols. BC and OC: black carbon and organic carbon.

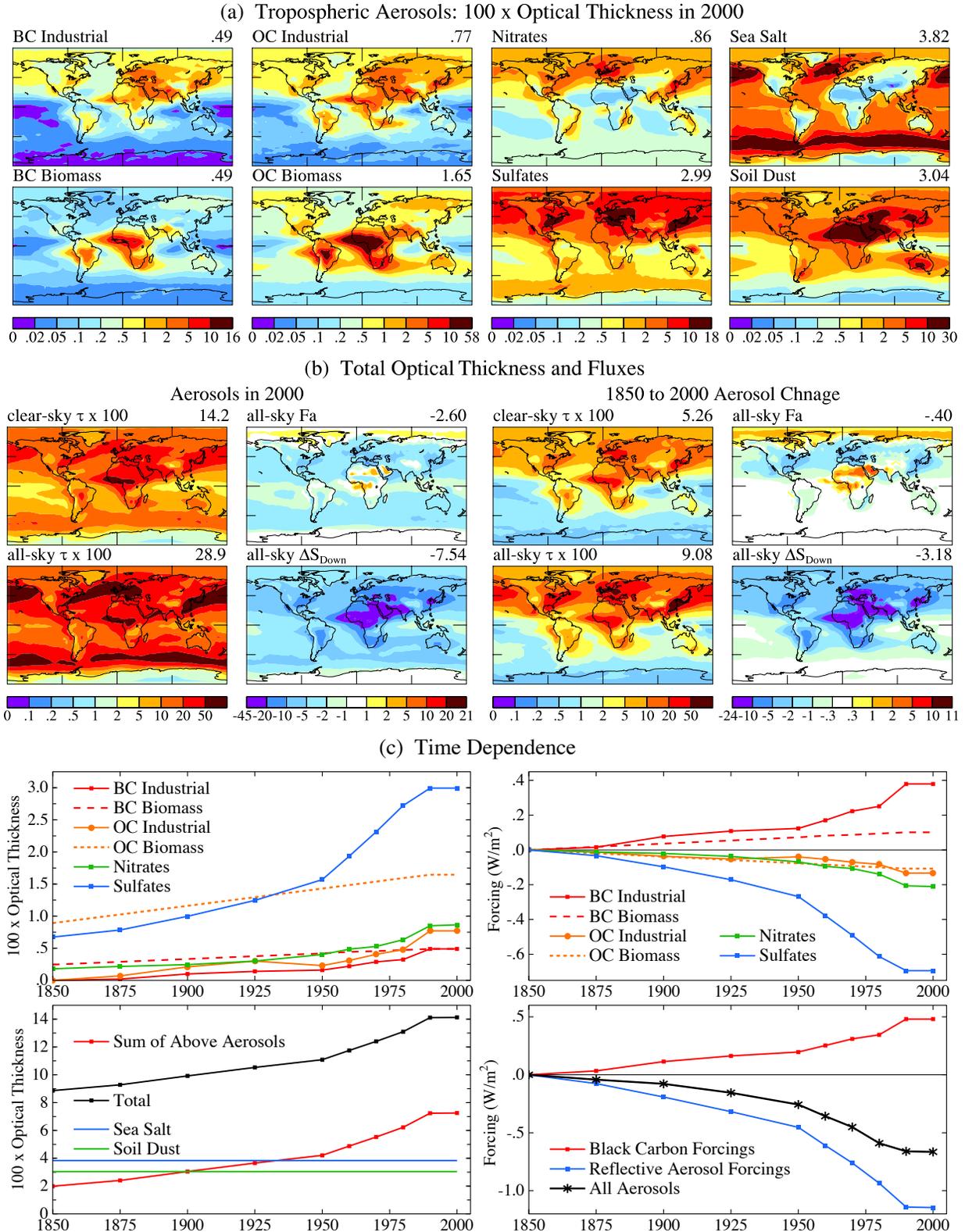

**Fig. 4.** The standard solar variability for most of our present climate simulations is from *Lean* (2000), which is based on irradiance observations since 1979 and solar proxy variables in prior years. The alternative solar forcing using only the Schwabe 11-year variability is from *Lean et al.* (2002).

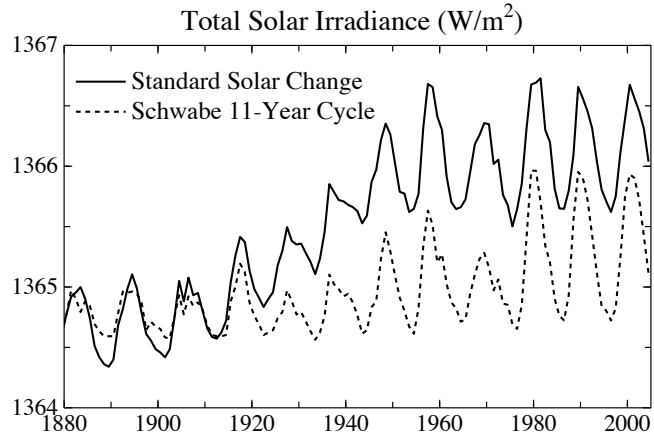

**Fig. 5.** Effective global climate forcings employed in our global climate simulations, relative to their values in 1880.

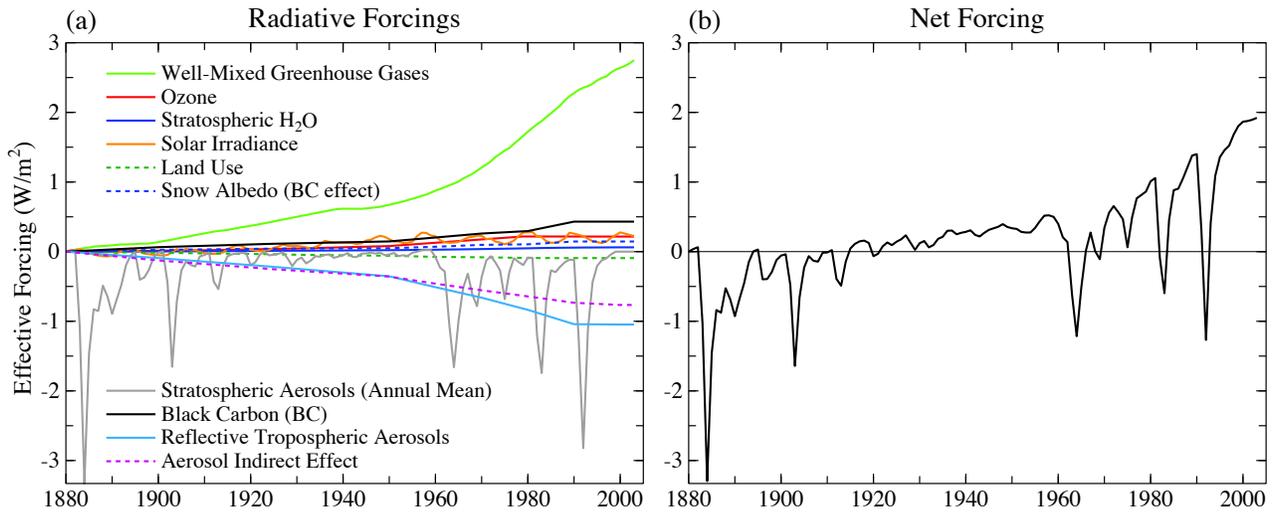

**Fig. 6.** Global mean climate change of GISS model III version of modelE atmosphere coupled to Russell ocean model and driven by the climate forcings of Fig. 5. Left column shows temperature change at three levels using vertical weighting functions of the MSU satellite instrument (*Mears et al.* 2003; see also www.ssmi.com/msu/msu_data_description.html). Observed surface temperature, in the upper right, uses meteorological station data over land and SSTs over the ocean. "Observed" ocean ice cover is the analysis of *Rayner et al.* (2003). A small radiation imbalance in the control run is subtracted from planetary net radiation.

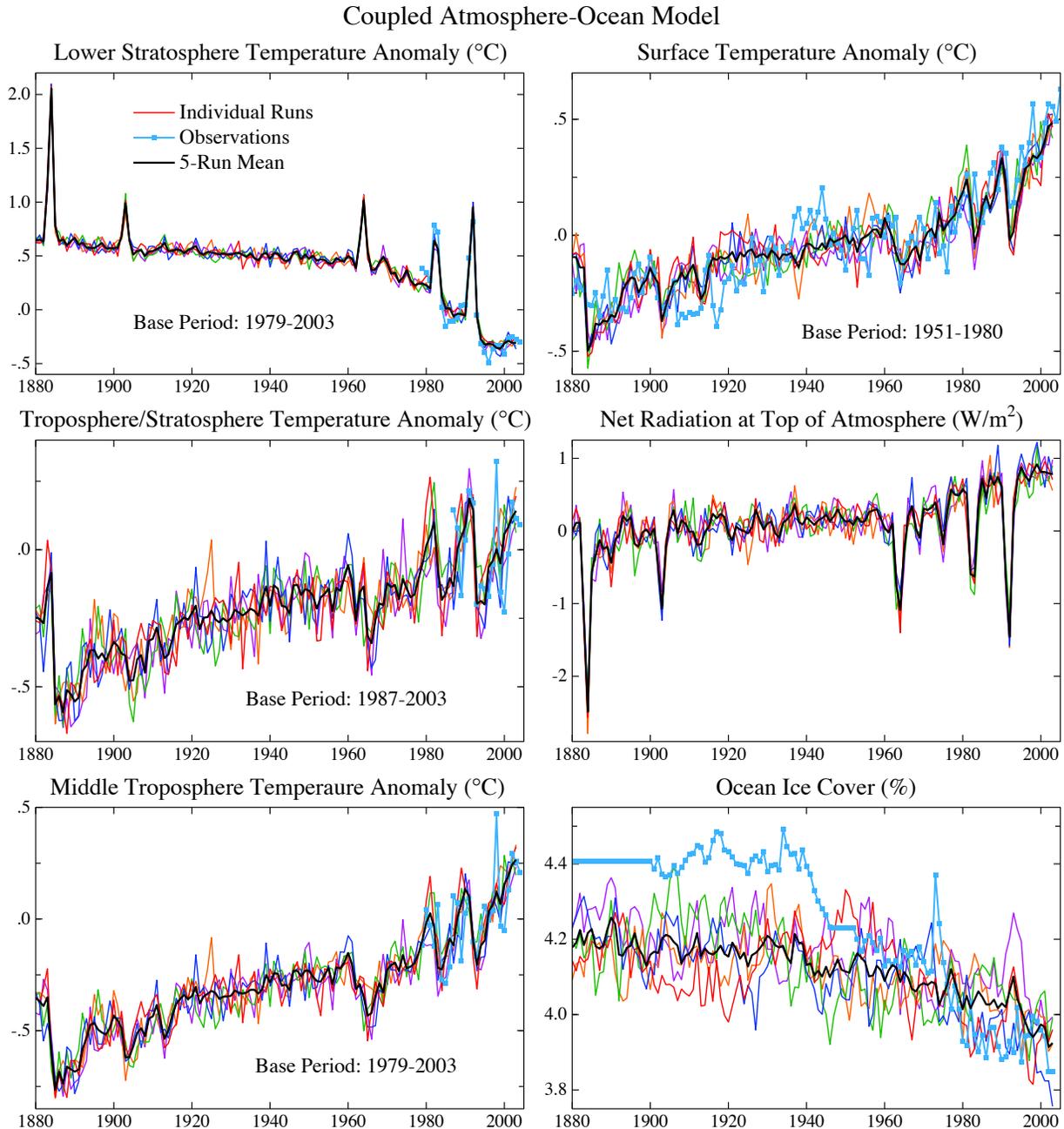

**Fig. 7.** Four of the six quantities in Fig. 6, but for alternative representations of the ocean: (a) ocean A, specified SST and sea ice of *Rayner et al.* (2003), and (b) ocean B, q-flux ocean (*Hansen et al.* 1984), with ocean horizontal heat flux inferred from control run of ocean A and locally varying diffusion of heat into 4 km deep ocean. As discussed in Sect. 5.3 and Supplementary Sect. S2, the sea ice change In observations between 1940 and 1945 is likely a spurious effect of data inhomogeneities.

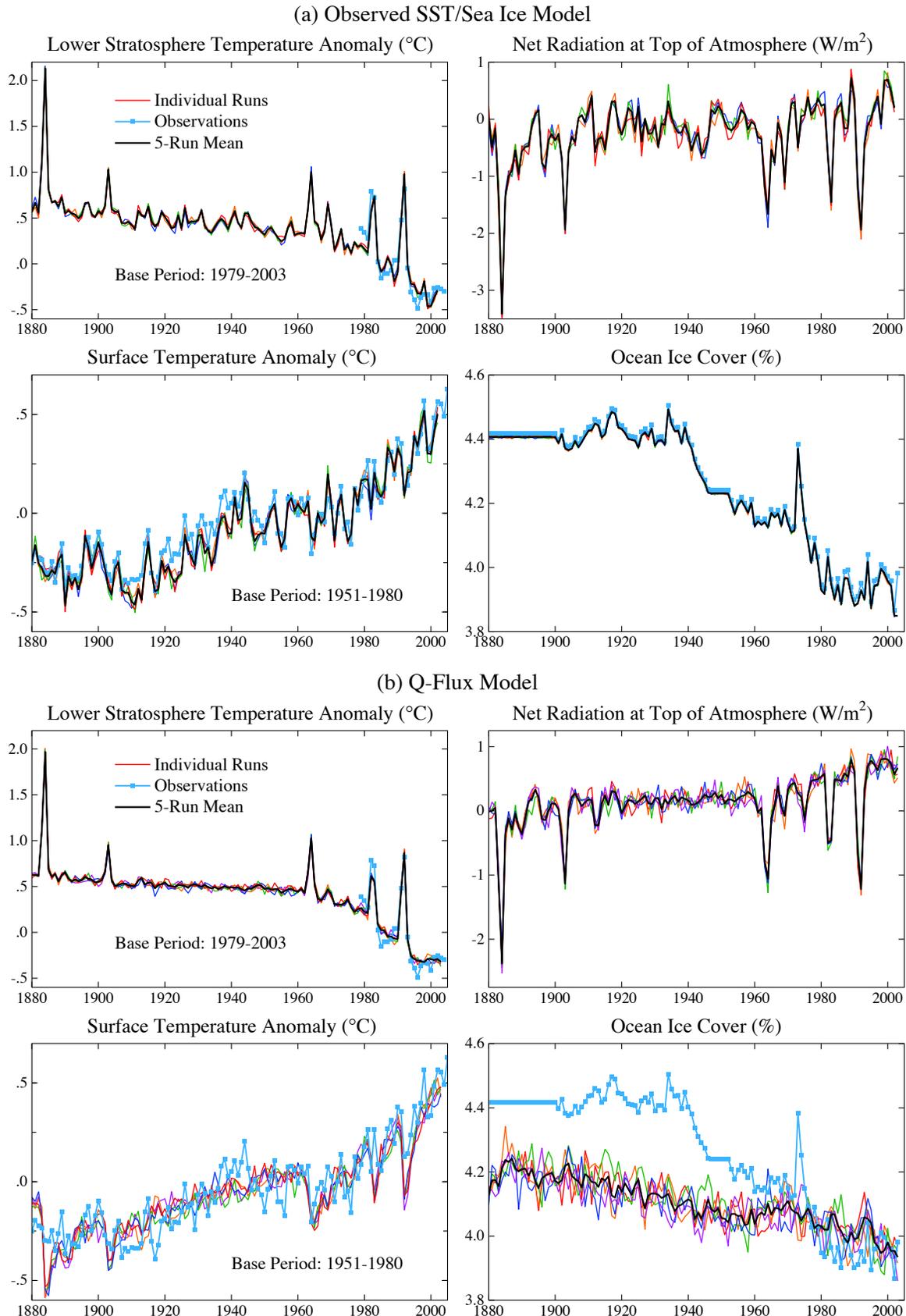

**Fig. 8.** Simulated global temperature response to all climate forcings of Fig. 5 operating at once, and the response to individual forcings. Five runs for each forcing were initiated at 25-year or 30-year intervals of the control run. Small model drift was removed by subtracting the control run year-by-year. Response to the aerosol indirect forcing is obtained as the difference between runs with direct and indirect forcings and runs with only the direct aerosol forcing. Climate model is the model III version of GISS modelE with the Russell ocean model.

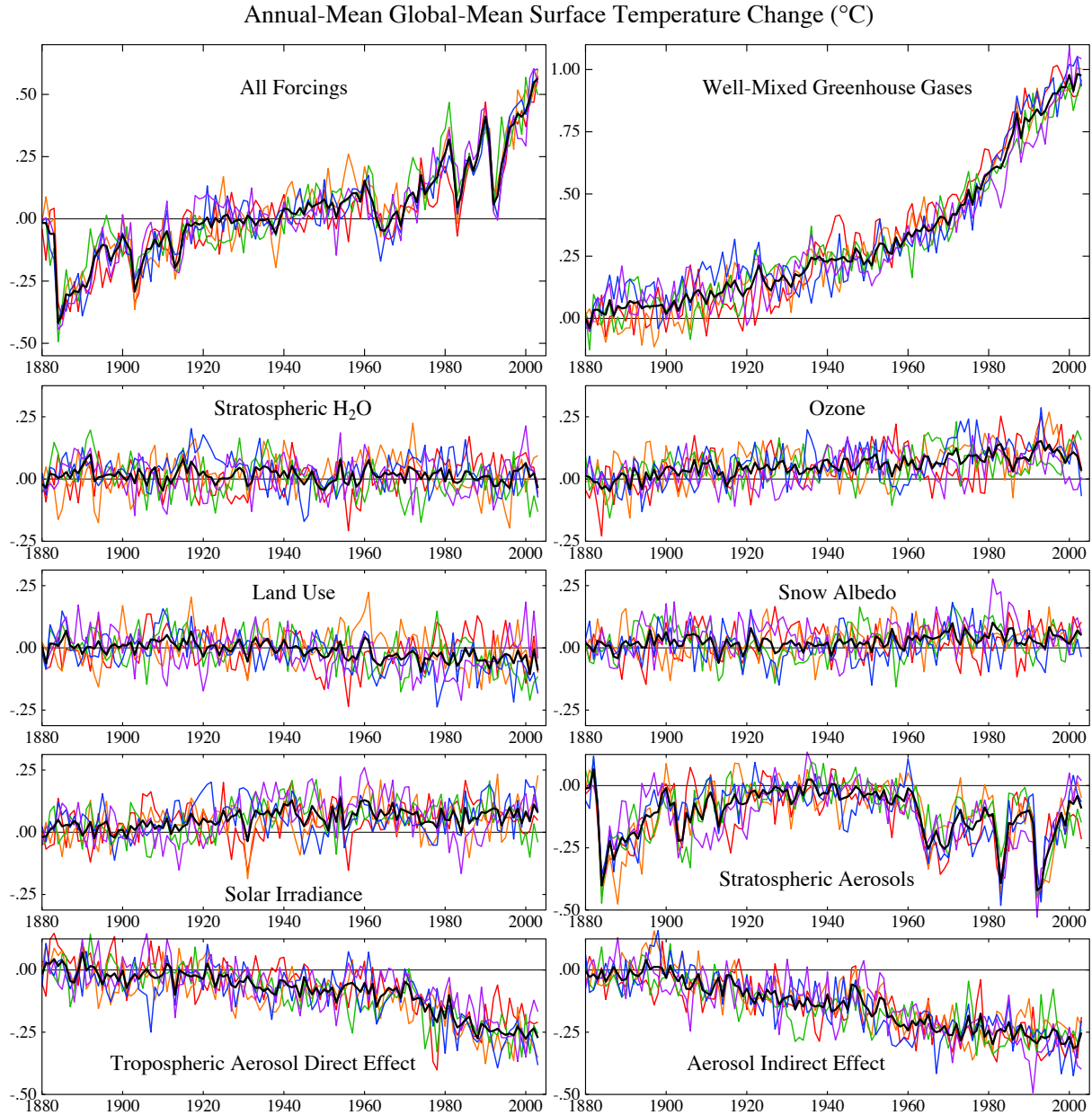

Annual-Mean Global-Mean Surface Temperature Change (°C)

**Fig. 9.** Global maps of temperature change in observations (top row) and in the model runs of Fig. 8, for 1880-2003 and several subperiods. Observations are based on analysis of *Hansen et al.* (2001), which uses urban-adjusted meteorological station measurements of surface air over land and SST data of *Rayner et al.* (2003) over the ocean.

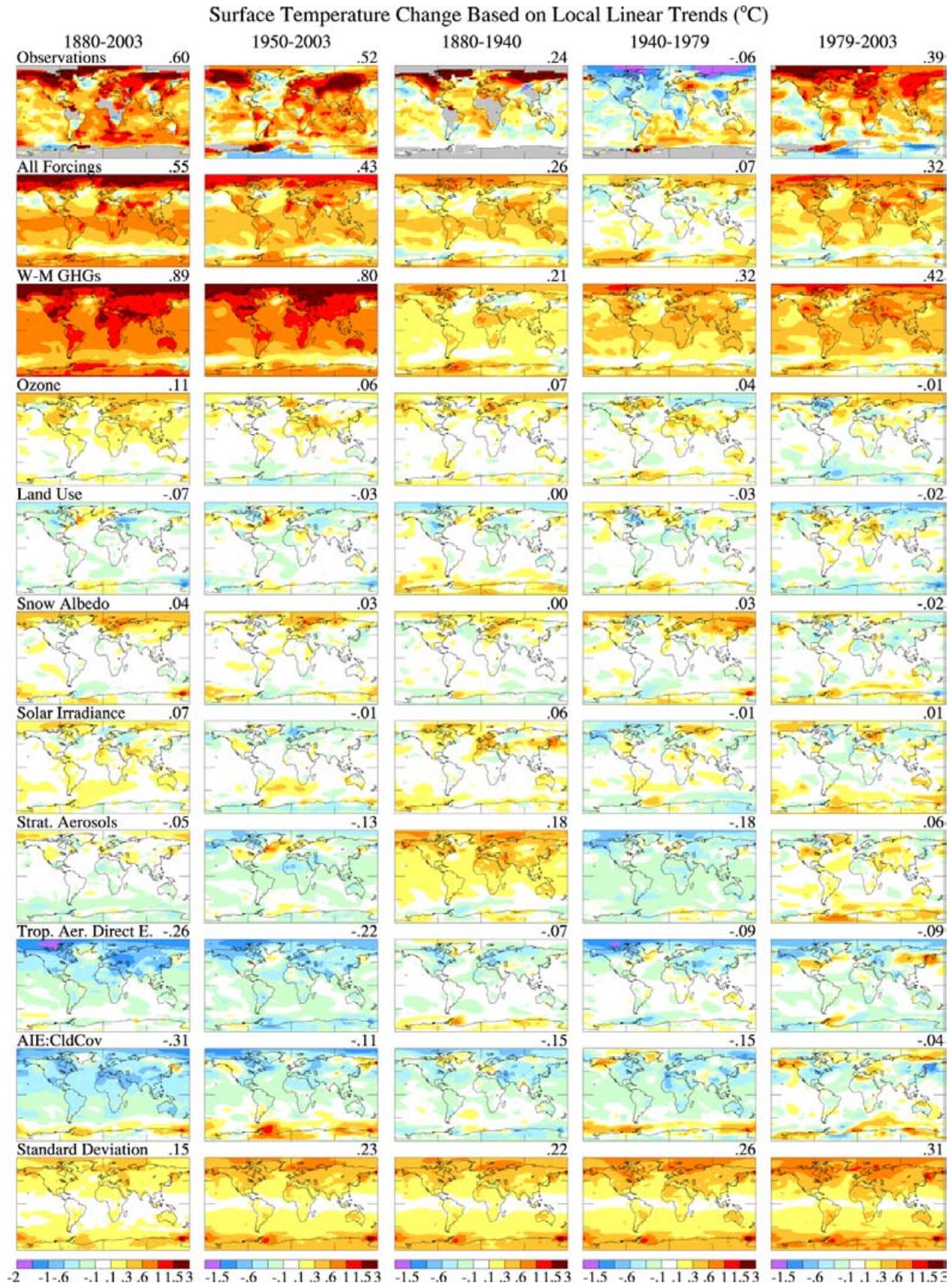

**Fig. 10.** Latitude-time annual-mean surface temperature anomalies relative to 1901-1930 for (a) observations and simulations driven by all forcings, (b) unforced variability in the control run (bottom two diagrams on left), and (c) simulations with individual forcings (right side).

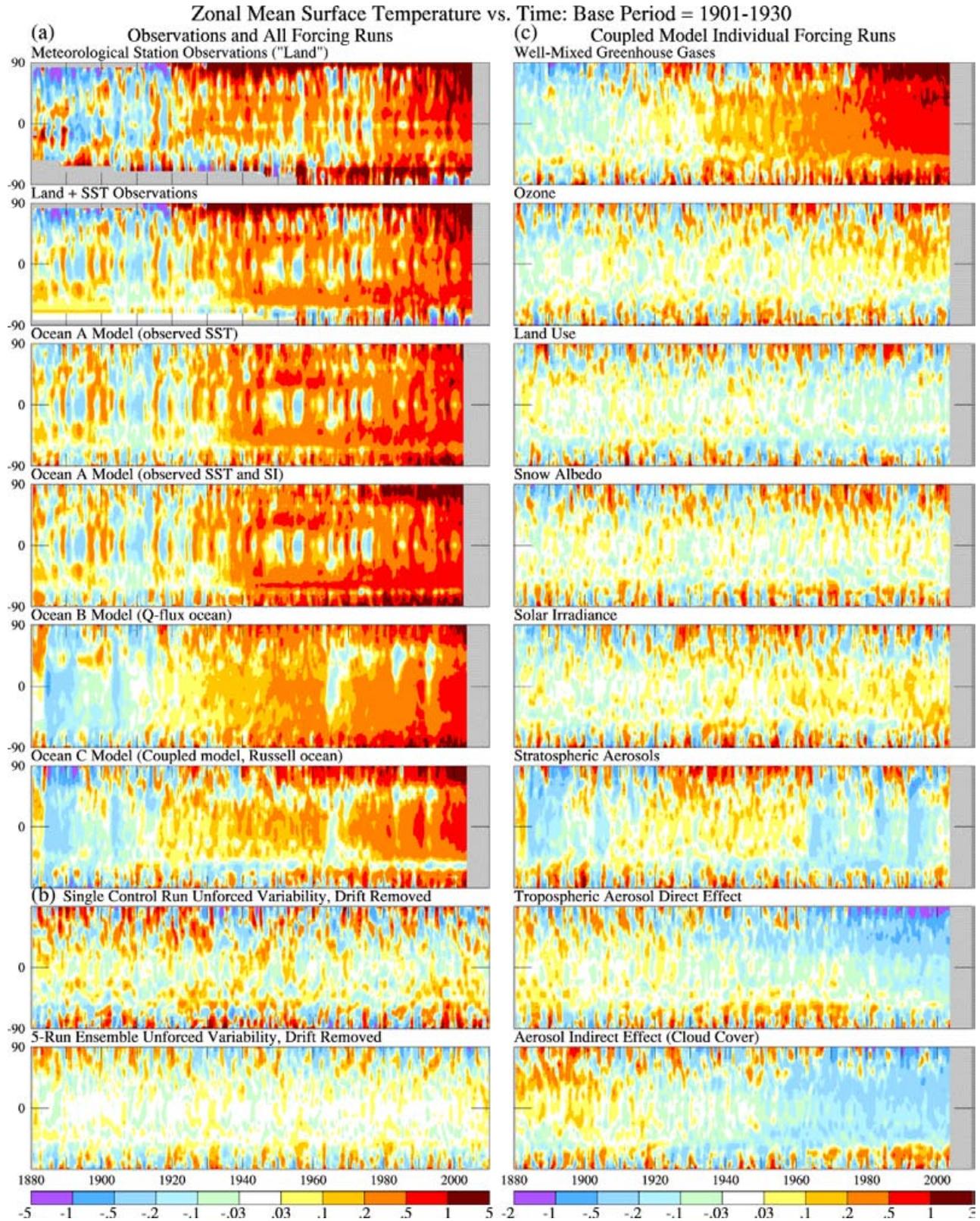

**Fig. 11.** (a) Ensemble-mean zonal-mean temperature change versus latitude and altitude (in hPa of pressure) for 1880-2003 period and three subperiods. Ensemble-mean changes > 2σ (lowest panel) are significant at > 99%. (b) Ensemble-mean zonal-mean temperature change versus month and latitude for the LS (lower stratosphere) level of microwave satellite observations (left two columns) and for surface air (right two columns). Changes > 2σ (lowest panel) are significant at > 99%.

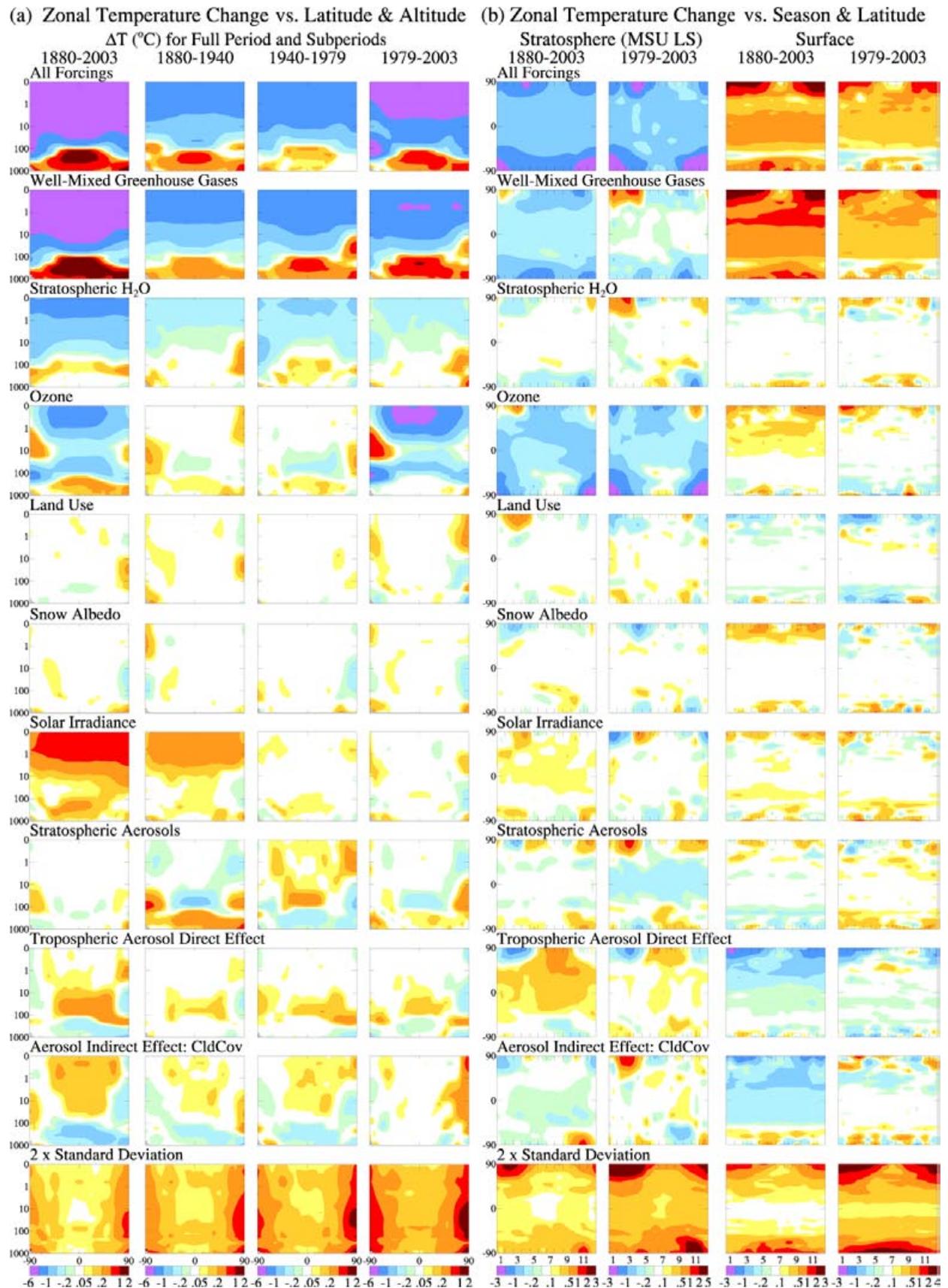

**Fig. 12.** Observed temperature change compared with simulations for the coupled climate model (ocean C) driven by all forcings of Fig. 5. (a) altitude-latitude temperature changes in three periods, for radiosonde data as graphed by *Hansen et al.* (2002) from analysis of *Parker et al.* (1997), and using satellite observations at three atmospheric levels (*Mears et al.* 2003; see also www.ssmi.com/msu/msu_data_description.html) and surface temperature (*Hansen et al.* 2001), (b) modeled tropopause height changes for all forcings and several individual forcings, (c, d) latitude-month temperature changes for levels observed by satellite and for the surface.

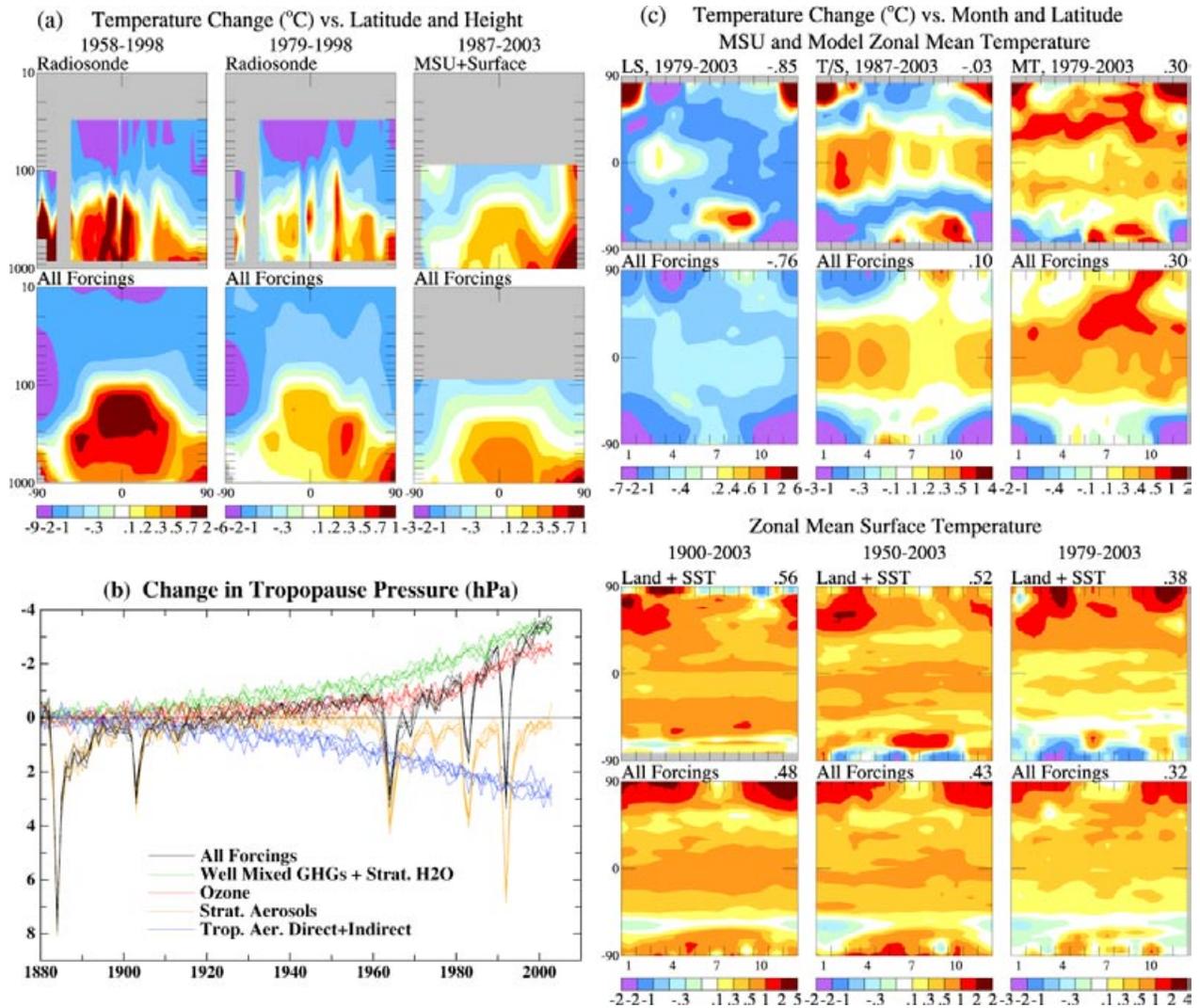

**Fig. 13.** Changes of indicated quantities simulated with ocean C for 1900-2003 based on trend of annual mean. Observed changes of several quantities are shown in the top row for specified periods. Numbers on upper right corners are global means; number on upper left corner of downward shortwave radiation is the area-weighted mean for gridboxes with observations. CRU: Climate Research Unit (*Mitchell et al.* 2004); GEBA: Global Energy Balance Archive (*Gilgen et al.* 1998).

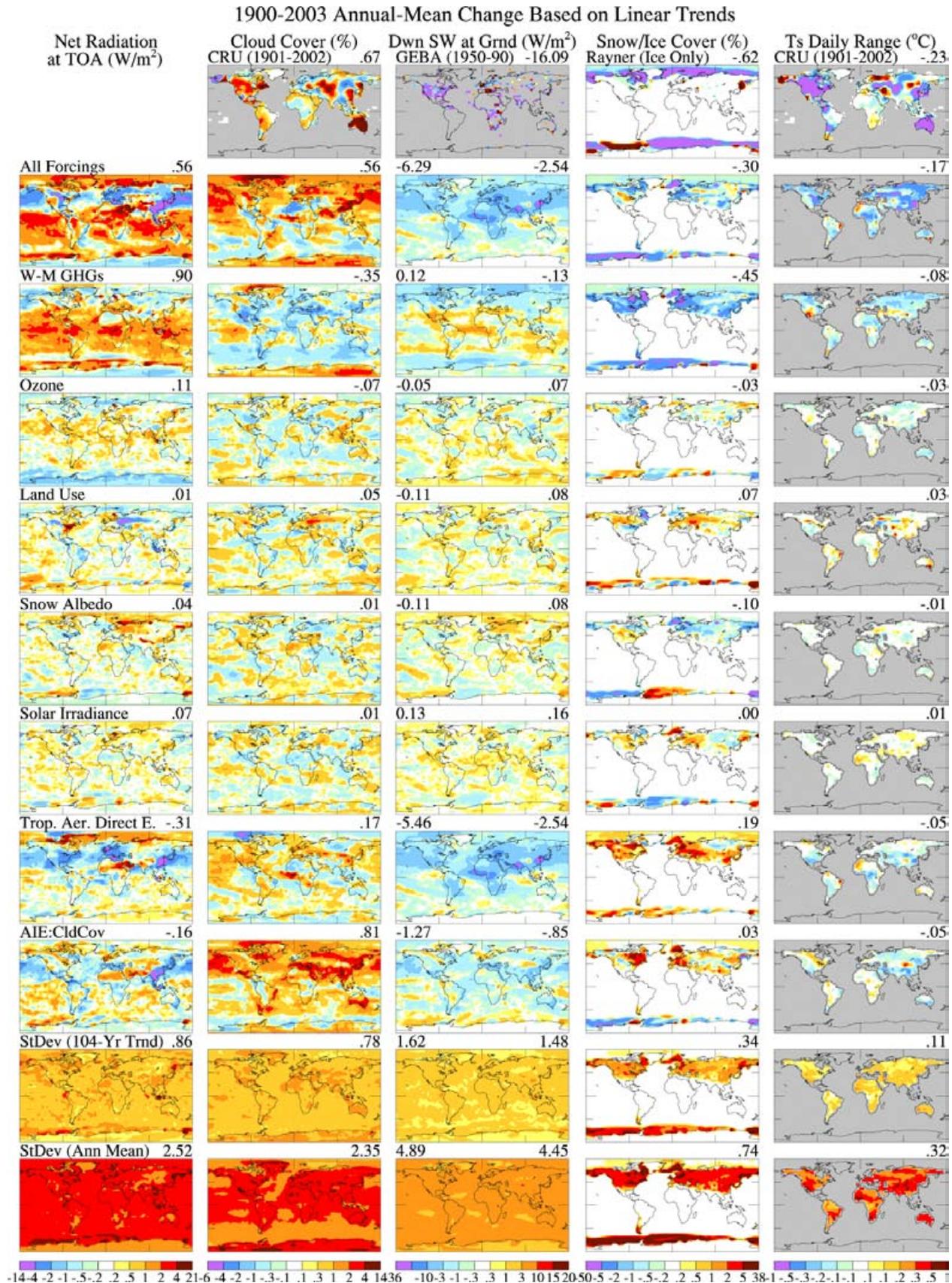

**Fig. 14.** Changes of indicated quantities simulated with ocean C for 1900-2003 based on trend of annual mean. Observed changes of several quantities are shown in the top row for specified periods. NCEP: National Center for Environmental Prediction reanalysis (*Kalnay et al.* 1996).

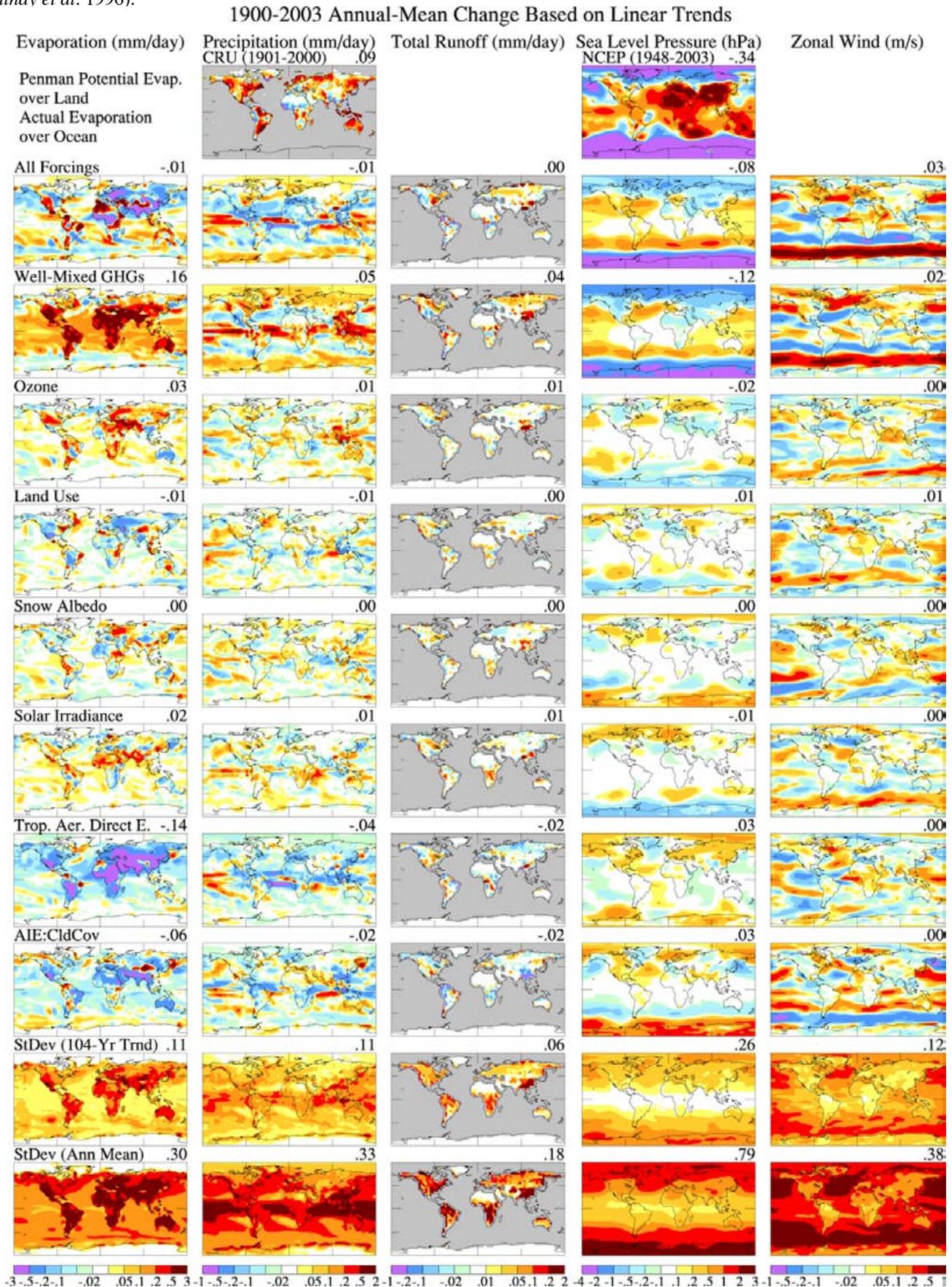

**Fig. 15.** Changes of annual-mean zonal-mean atmospheric temperature, water vapor, zonal wind, and stream function versus latitude and altitude (in hPa of pressure) for ocean C simulations based on linear trends over periods (a) 1950-2003 and (b) 1979-2003.

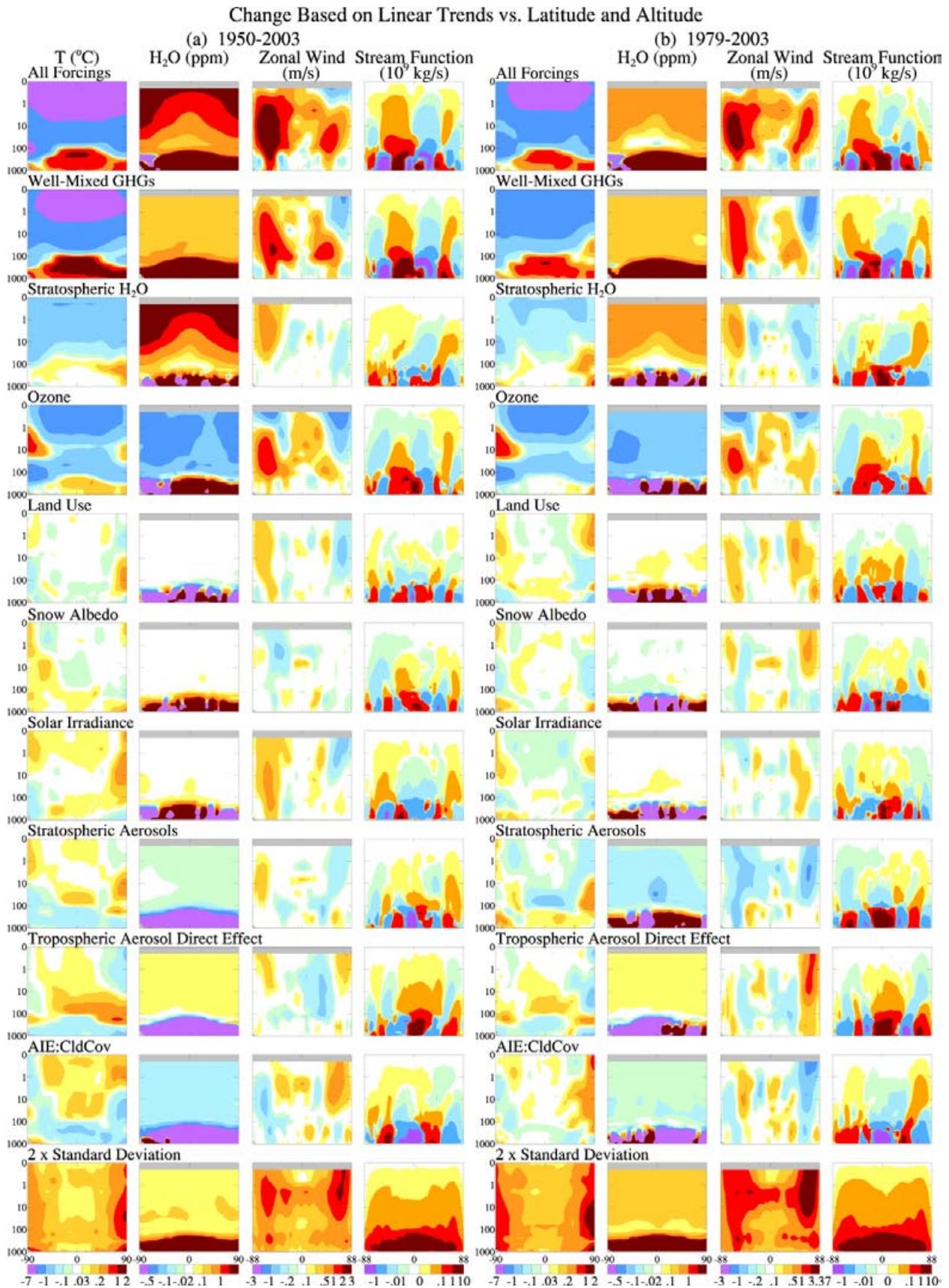

**Fig. 16.** Surface temperature change in observations and simulations for standard 'all forcings' scenario, two aerosol sensitivity runs [½ ΔSulfate and ½ ΔSulfate + 2 × Δ (biomass burning BC and OC)], and the alternative solar forcing, AltSol. AltSol includes only the Schwabe 11-year solar variability of *Lean et al.* (2002). Gray area is region within one standard deviation for 5-run ensemble with standard "all forcings".

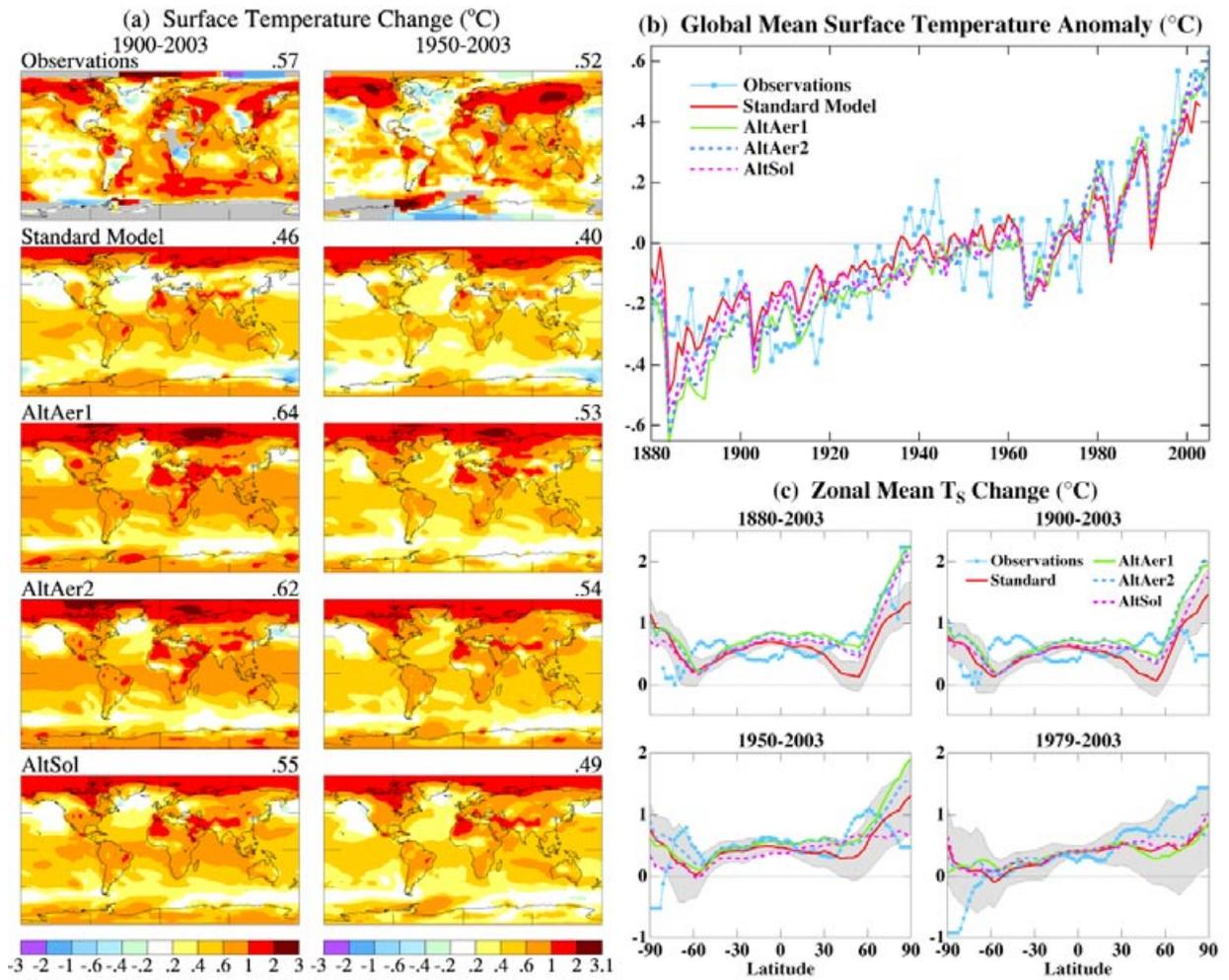

**Electronic Supplementary Material**

**for**

**Climate simulations for 1880-2003 with GISS modelE**

**by J. Hansen et al.**

## S1. Alternative Data Samplings and the Krakatau Problem

Comparisons of simulated climate and observations commonly involve choices that influence how well the model and data appear to agree. Choices of surface temperature data deserve scrutiny, because surface temperature provides the usual measure of long-term 'global warming' as well as a test of possible global cooling after large volcanic eruptions. We illustrate here alternative comparisons of model and observations, with model results being those of the coupled model (ocean C, the *Russell et al.* [1995] model) driven by all climate forcings of Fig. 5. This model run is discussed in later sections.

### S1.1. Century time-scale

Temperature measurements at meteorological stations provide a reasonably consistent data set for continental and island locations (*Jones et al.* 1986; *Hansen and Lebedeff* 1987), albeit one in which the station records are spatially inhomogeneous, often broken temporally, and subject to non-climatic effects. The meteorological station records that we employ have been adjusted for urban effects using neighboring rural stations (*Hansen et al.* 2001). Such adjustments are imperfect, but the impact on global mean 100-year temperature change of uncertainties in urban adjustments is not larger than about 0.1°C (*IPCC* 2001). For the short interval after a volcano considered here, urban adjustments are negligible.

The GISS analysis of station data (*Hansen et al.* 1981; *Hansen and Lebedeff* 1987) combines stations with overlapping periods of record so as to preserve information on temporal variability while allowing individual stations to affect estimated temperature change at distances up to 1200 km. It has been shown, by sampling a global model with realistic temperature variability at the station locations, that after about 1880 the station network is capable of yielding a good estimate of global temperature change despite poor coverage in the Southern Hemisphere (*Hansen and Lebedeff* 1987). However, island and coastal stations fail to sample part of the ocean, and both observations and models indicate that the long-term temperature response tends to be less over the ocean than over continental areas. Thus we expect the long-term "global" temperature change estimated from the meteorological station network alone to slightly overestimate true global mean temperature change.

Improved global coverage is obtained by combining meteorological station data with sea surface temperature (SST) measurements of ocean areas (*Jones et al.* 1999; *Hansen et al.* 2001). However, ocean data have their own problems, including changes of measurement methods and infrequent sampling of large regions (*Parker et al.* 1994). Sampling is especially poor in the 1800s, and spatial-temporal data-fill procedures risk smoothing real variations. In addition, the ocean area with the largest response to climate forcings in climate models, regions of sea ice, are practically unobserved.

Fig. S1a shows an estimate of global temperature using only meteorological station measurements (*Hansen et al.* 2001). The observed 1880-2003 temperature change, based on the linear trend, is 0.69°C in this case. The model 5-run "all forcings" ensemble mean yields 0.56°C, with the model result being a true global mean.

Fig. S1b uses the same land temperatures as in Fig. S1a, but it adds SST data for the oceans, using ship data of *Rayner et al.* (2003) for 1880-1981 and subsequently satellite data (*Reynolds and Smith* 1994; *Smith and Reynolds* 2004). Inclusion of ocean data reduces the observed global temperature change to 0.59°C. It also practically eliminates evidence for cooling after the 1883 Krakatau eruption.

Fig. S1c is a third alternative, comparing observed temperature from meteorological stations with the model sampled at the places and times of the observations. This sampled model data is run through the same temperature analysis program as observations to produce the global mean. This third procedure is optimum in the sense of having the most consistent treatment of model and data, as well as preserving a best estimate for high frequency temperature change in the period of sparse observations in the 1800s. The model sampled at observing stations yields a global warming of 0.59°C based on the linear trend, which is less than the observed 0.69°C. This discrepancy occurs because the model warms less over land areas than observed, a result that we identify with excessive anthropogenic tropospheric aerosols over Eurasia in our standard "all forcings", as discussed in Sect. 5. This third procedure provides a clean comparison of model and observations, but the integration over the globe is not a true global mean. In addition, it is unlikely that most modelers will sample their model results at the times and places of meteorological station measurements and run the results through the GISS temperature analysis program, thus making it difficult to compare GISS model results with other models.

Fig. S1d is a fourth alternative, comparing model results for the true global mean with observations that use only meteorological stations for 1880-1900 but add ocean data for 1900-2003, when ship data had better coverage. This alternative preserves temperature variations in the 1800s without exaggerating long-term global temperature change. The observed 1880-2003 temperature change in this case, 0.61°C, is slightly larger than in Fig. S1b, as expected due to the cooling in the 1880s. The disadvantage of this fourth alternative is the arbitrariness inherent in concatenating two data sets.

We present all four alternatives to help readers make their own assessment. For simplicity we use the procedure of Fig. S1b in following sections, i.e., we use the true global mean for the model and the land + ocean data for observations. However, it should be born in mind that these observations probably miss some actual cooling after Krakatau.

We examine the Krakatau period in more detail, because it has an effect on how well the model and observations appear to agree over the 120-year temperature record. We find it useful to compare the 1883 Krakatau and 1991 Pinatubo eruptions, the two largest volcanic aerosol climate forcings in the period of instrumental climate data (*Sato et al.* 1993). These volcanoes have the best chance of producing signals above the climate noise level and the Pinatubo period has extensive climate observations.

### S1.2. Temperature change after Krakatau and Pinatubo

Estimated aerosol optical depths after Krakatau and Pintubo are shown in Fig. S2a. The shape of the Krakatau curve is assumed to be similar to that after Pinatubo, as they were both low latitude injections to high altitudes at similar times of year. Measurements of decreased solar irradiance integrated over three years after Krakatau were used to set the aerosol optical depth (*Sato et al.* 1993). Effective forcings are shown in Fig. S2b. Resulting temperature anomalies, relative to the three-year mean preceding the eruption, are shown in Fig. S2c. The simulated cooling after Krakatau exceeds that after Pinatubo by more than the assumed 10% difference in their forcings. This must be at least in part because of planetary radiation imbalance of about +0.5 W/m$^2$ that existed just prior to the Pinatubo eruption (*Hansen et al.* 1993) but not at the time of the Krakatau eruption. Further, as mentioned in Sect. 3.2.1, the response to the Krakatau aerosols would have been reduced about 10 percent if the control run ocean temperatures had included the effect of prior volcanic eruptions via a mean stratospheric aerosol optical depth.

Fig. S2c shows that the global mean temperature based on meteorological station data after Krakatau is consistent with the climate simulations. The seasonal mean 1σ error bar for global temperature estimated from the meteorological station network in the 1880s is 0.15°C (*Hansen and Lebedeff* 1987). Thus the cooling observed by the station network after Krakatau for a given season could be a sampling error, but not the nearly continuous cool period for several years after the eruption. Furthermore, comparison of the global temperature curve estimated from meteorological stations in the Pinatubo era (right side of Fig. S2c) with the global temperature curve that has complete ocean coverage from satellite data shows that the station network tracks the complete global data within the expected error for the station network (1σ sampling error being 0.09°C for the station distribution in the 1990s). We conclude that there was global cooling after Krakatau.

Fig. S2d shows the observed and simulated surface temperature anomalies in the northern winter (DJF) following the eruptions and the northern summer (JJA) about one year after the eruptions. As expected, the model and observations show strong cooling in the summer after the eruption, especially over the continents. Also, the model and observations show global cooling in DJF, with evidence for regional Eurasian "winter warming", an expected dynamical response (*Groisman* 1992; *Perlwitz and Graf* 1995; *Robock* 2000; *Shindell et al.* 2001), which has previously been reported to occur in current GISS models (*Shindell et al.* 2004). The model, using the coarse-resolution *Russell et al.* (1995) ocean, is not able to produce El Ninos, which have accompanied several large volcanoes in the past century (*Handler* 1984; *Robock* 2000; *Mann et al.* 2005) and may be responsible for warming in the region of Alaska. Temperature anomalies are muted in the 5-run model mean in Fig. S2d, but the magnitude of anomalies is more realistic in the individual runs, which are available on the GISS web-site.

## S2. Mean Stratospheric Aerosols in Control Run

Our control run had no stratospheric aerosols. Aerosols from the 1883 Krakatau eruption caused ocean heat content in the experiment runs to fall below that in the control run, as expected. However, despite steadily increasing greenhouse gases, the ocean heat content did not recover to that of the control run until about 2000. In reality, ocean temperature is also influenced by volcanoes that erupted prior to 1880. Ideally, ocean initial conditions in 1880 would be obtained from a spin-up run that had time-dependent forcings, including volcanoes, for several centuries prior to 1880. That is not usually practical, if for no other reason than the absence of information on earlier volcanic eruptions. However, it is easy to include a mean stratospheric aerosol amount in the control run.

Current control runs with our model include a mean stratospheric aerosol optical thickness τ = 0.0125 at 0.55 μm wavelength, which is the 1850-2000 mean value of the *Sato et al.* (1993) aerosol climatology. The equilibrium global (surface) cooling for τ = 0.0125 (10% of the maximum τ for Pinatubo) is ~0.2°C, and the effect on deeper ocean temperatures is sufficient to alter the rate of ocean heat storage in transient climate simulations. Using a control run in which the ocean temperature had equilibrated with an atmosphere including this mean aerosol amount, we carried out an ensemble of runs for 1850-2003. The concentration of volcanoes near the end of the 19th century caused the ocean heat content anomaly to be negative for several decades, but it recovered to the control run value by the mid 20th century and it subsequently increased at a rate comparable to that reported by *Levitus et al.* (2000).

## S3. Control Run Disequilibrium and Drift.

Our coupled atmosphere-ocean (ocean C) simulations, to meet the deadline for submission to IPCC, were initiated before the control run (which provides initial conditions for the experiments) had reached equilibrium, i.e., while there was still an imbalance between the amounts of energy absorbed and emitted by the planet. As a result, the model response to any forcing included a small drift.

We minimize drift effects by subtracting, year-by-year, the same quantities from the same period of the control run. This procedure yields diagnostics with 'double noise', i.e., it contains unforced variability of both the control and experiment runs, while the real world has only a single source of unforced variability. Double noise can be minimized by initiating additional control runs at the same points at which experiments are initiated.

An alternative way to remove drift is to calculate and subtract from the experiment result the mean drift in the control over the period of the experiment. For example, for a 124-year 1880-2003 experiment initiated at year X of the control run, we could calculate the linear trends of control run diagnostics over the period X to X + 123 and subtract the control run diagnostics based on their linear trends from the corresponding quantities in the experiment run. This alternative procedure avoids year-to-year double noise, but it does not eliminate drift effects entirely because variability occurs on all time scales.

Noise effects were exacerbated by the fact that most of our experiments, with individual forcings and with multiple forcings, were initiated at the same points of the control run. The control run has unforced variability not only interannually, but on 124-year and all other time scales. Thus when we add up responses to individual forcings, with drift subtracted, we are including the same unforced 124 year fluctuation for each forcing. Therefore we cannot expect the sum of the responses to individual forcings to equal the response to the sum of the forcings, even if there is no non-linearity in the climate response.

An improved procedure would be to initiate experiments for different forcings at different points on the control run, in addition to spacing ensemble members. It would perhaps be still better to carry out a long control run that reaches equilibrium before experiments are initiated, so there would be no need to subtract a control run. However, the merits of waiting until the control run equilibrates before initiating experiments may be reduced if the equilibrium climate drifts too far from the real world.

**S4. Surface Temperature Definition.**

Surface air temperature (Ts) in modelE is calculated at 10 m height. The land-ocean temperature index (Tx) (*Hansen et al. 2001*) is from observations at 2 m height at meteorological stations and SST data of *Rayner et al.* (2003) and *Reynolds and Smith* (1994) over the ocean. Temperature *changes* of model and observations are compared, which minimizes, but does not eliminate, the effect of these height differences.

Fig. S3 shows the modeled 1880-2003 temperature change for (1) ocean A driven by no forcings except SST and sea ice change, (2) the same as (1) but including "all forcings" (GHGs, aerosols, etc.), (3) the same as (2), but for the coupled atmosphere-ocean climate model. For each of these three models we show the global surface air temperature (Ts), the temperature index (Tx), which uses the ocean temperature instead of Ts for ocean areas, and their difference.

In the case of ocean A with no forcings, Ts and Tx are practically the same on global average, even though there are regions where they differ by a few tenths of a degree. In the case of ocean A with radiative forcings, the forcings are able to change atmospheric temperature slightly even though SST is fixed. Global mean Ts increases 0.03ºC more than Tx increases over the period 1880-2003. In the case of ocean C the ocean temperature is able to respond to the change of near surface temperature gradient, and Ts increases 0.05ºC more than Tx increases over the 1880-2003 period.

These comparisons indicate that our use of global Ts at 10 m height overstates global mean $\Delta T$ by several hundredths of a degree, if our aim is comparison with a temperature index that uses SSTs. We could employ $\Delta$Tx from the model based on the first layer ocean temperature, but that would be inconsistent with the procedure used in previous studies with the GISS model and other models, and thus we used $\Delta$Ts in this paper. This issue may be noticeable only in the GISS model, which calculates Ts in an iterative fashion (*Hansen et al.* 1983; *modelE* 2006). In the future the issue might be practically eliminated by calculating Ts at 2 m height, rather than 10 m.

These small changes in $\Delta T$ do not alter the geographical pattern of the discrepancy between model and observations. The main implication is that the 124-year warming in our model with "all forcings" is ~0.10ºC less than observed, rather than 0.05ºC less. Thus the need for less tropospheric aerosol amount becomes clearer in the global mean temperature, as well as from unrealistic cooling over Europe.

As future models are better able to simulate observed climate change, it will be worth removing any such discrepancy in comparison with observed surface temperature. We are uncertain whether this comparison issue exists for other climate models.

**S5. Ozone Scenario.**

The first set of runs that we provided to IPCC inadvertently used the *Randel and Wu* (1999) decadal rate of stratospheric $O_3$ depletion as the 18-year change, thus understating stratospheric $O_3$ depletion by the factor 10/18. Corrected runs were submitted several months later, and both sets of runs remain available at www-pcmdi.llnl.gov/ipcc/about_ipcc.php. The correction reduced the 1880-2003 global forcing Fa by 0.03 W/m². The main impact of the correction was on stratospheric cooling in the Antarctic region during the time of $O_3$ depletion, with the corrected results providing better agreement with observations. The present paper and *Efficacy* (2005) use the corrected $O_3$ change.

A second issue with the $O_3$ scenario concerns $O_3$ forcing due to tropospheric pollution. The $O_3$ scenario was derived from an off-line simulation of a tropospheric chemistry model (*Shindell et al.* 2003), which yielded an 1880-2000 $O_3$ change from the surface to the 150 hPa level at all latitudes. Global forcings for this $O_3$ change were Fi = 0.44, Fa = 0.38 W/m². However, tropospheric $O_3$ forcing implemented in our transient simulations was less, as high-latitude $O_3$ increases above the model's tropopause (Fig. 3 of *Efficacy* [2005]) were excluded, reducing $O_3$ forcing by 0.05 W/m². As the pollution effect on $O_3$ at low latitudes was only allowed to reach the 150 hPa level, we suspect that our total $O_3$ forcing (Fa = 0.28, Fs = 0.26, Fe = 0.23 W/m², including tropospheric pollution and stratospheric depletion, from Table 1) underestimates actual $O_3$ forcing. Future $O_3$ scenarios should be generated by models with improved vertical resolution and higher model top, preferably integrating effects of tropospheric pollution and stratospheric change.

**S6. Snow Albedo.**

A computer programming error was present in the calculation of snow albedo in several of our climate simulations. Some of these runs were repeated with the error corrected, as delineated below. Our intention was for snow albedo change to be proportional to BC deposition as calculated by the aerosol transport model of *Koch* (2001). The error caused albedo change to be exaggerated in partially snow-covered land gridboxes and understated over sea ice, because total albedo change was fixed.

Our initial 'all forcing' run provided to IPCC contained both the ozone error (A.4) and snow albedo error (A.5). We also provided to IPCC 'all forcing' runs with the ozone error corrected and later runs with both errors corrected. Because of space limitations, the DOE web site includes only the original 'all forcing' ensemble and the ensemble with the ozone forcing corrected. All three ensembles are available on the GISS web site.

The 'all forcing' and snow albedo alone ensembles were rerun with the snow albedo error corrected. The corrected program was also used in 'Arctic pollution' runs (Fig. 5 in *Dangerous* [2006]). However, the AltAer1, AltAer3, and AltSol runs contain the snow albedo error, but not the ozone error. These ensembles were not rerun with corrected snow albedo because of the small magnitude of the error and the fact that it would not alter conclusions from those runs. To allow precise comparison with AltAer1, AltAer2, and AltSol, the standard model results in Fig. 16 of this paper and Fig. 6 in *Dangerous* (2006) are the 'all forcing' results that include the snow albedo error.

The simulations employed in the energy imbalance study of *Hansen et al.* (2005b) contained both errors. The errors in global forcing, +0.03 W/m² and –0.02 W/m², opposed each other, but regional and temporal effects would not cancel, e.g., stratospheric cooling over Antarctica was underestimated. However, the magnitude of these errors is too small to affect conclusions of that paper.

*Efficacy* (2005) simulations included the snow albedo error but not the ozone error. In Table 4 and Fig. 16 of *Efficacy* (2005) the snow albedo forcing was calculated with the incorrect program. Fa was actually 0.05 W/m², not 0.08 W/m², and the correct efficacy for the snow albedo effect was Ea ~ 2.7, not Ea ~ 1.7.

**References to Supplementary Material**

**Supplementary Figures**

**Fig. S1.** Observed and modeled global surface temperature change for alternative ways of averaging over the globe, with the model driven by all forcings of Fig. 5. (a) Observations are surface air temperature at meteorological stations averaged as defined by *Hansen et al.* (1999), model is true global mean. (b) Observed temperatures are surface air measurements at meteorological stations combined with SST measurements over the ocean, model is true global mean. (c) Observations are at meteorological stations as in Fig. S1a, model is sampled at the same places and times and analyzed in the same way as observations. (d) Model is true global mean, observations are based only on meteorological stations during 1880-1900, but incorporate SSTs after 1900.

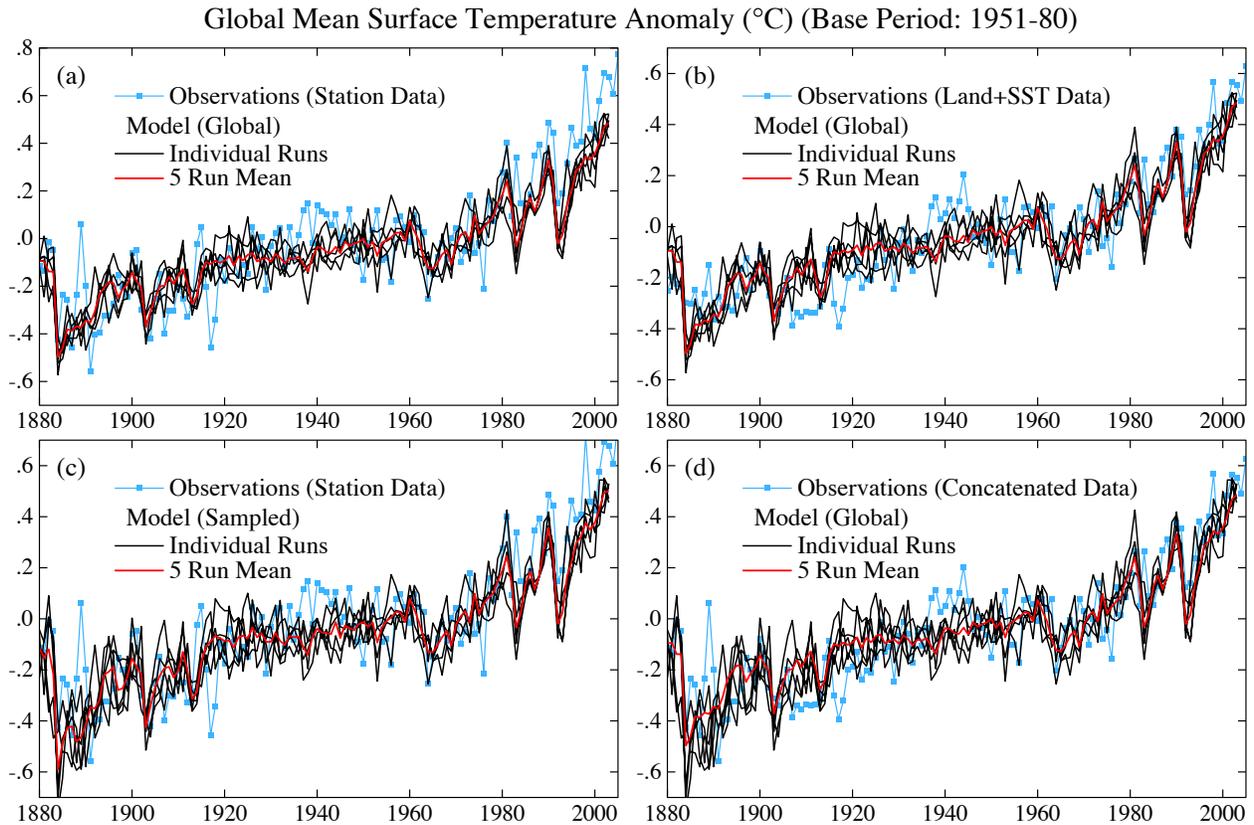

**Fig. S2.** (a) Stratospheric aerosol optical thickness and (b) effective forcing for the assumed aerosol scenario, based on update of *Sato et al.* (1993). (c, d) Temperatures simulated by the climate model normalized to the mean for the 36 months before the eruption, with the circles and asterisks in (c) being the Jun-Jul-Aug and Dec-Jan-Feb means, respectively. Observed 'station' data and 'land + ocean' are based on analyses of *Hansen et al.* (2001), using, respectively, meteorological stations alone and those same stations plus ocean data of *Rayner et al.* (2003).

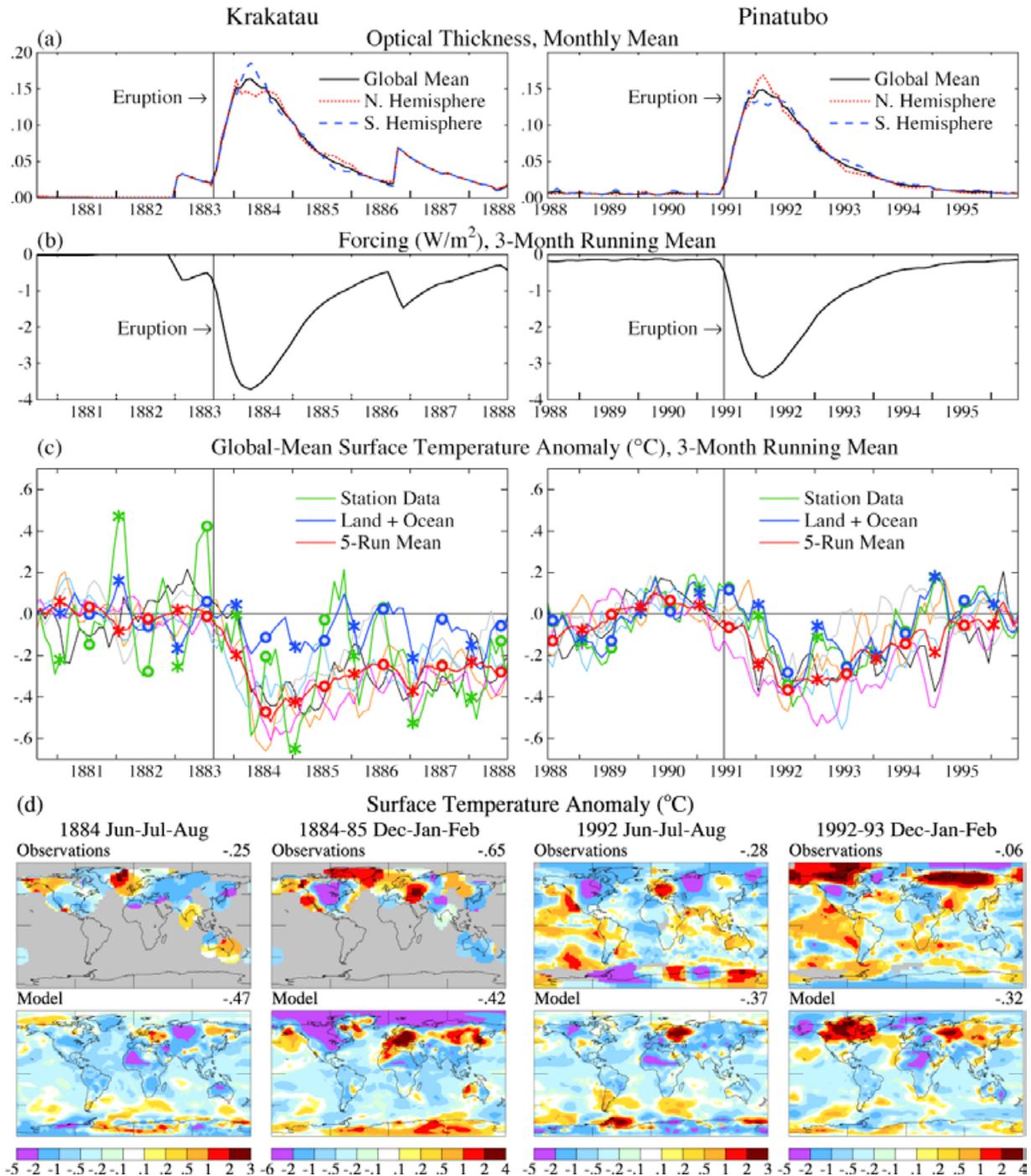

**Fig. S3.** Simulated surface temperature change for 1880-2003 based on local linear trends. Ts is the surface air temperature at 10 m altitude, Tx substitutes SST for Ts over the ocean. Ocean A uses the SST and sea ice history of *Rayner et al.* (2003) coupled to atmospheric modelE, while ocean C couples modelE with the *Russell et al.* (1995) ocean model.

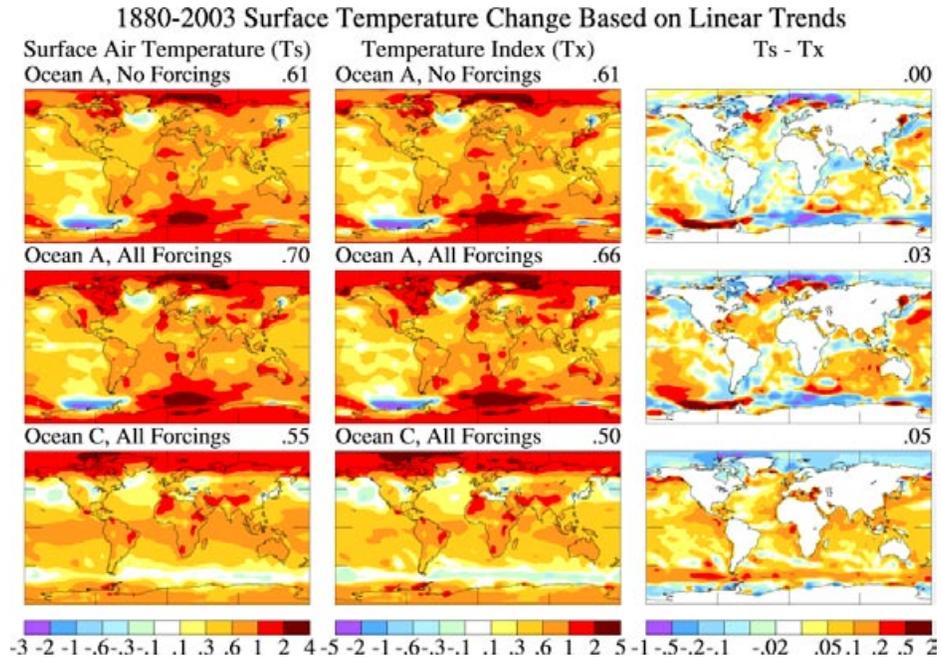